\documentclass[10pt,twocolumn,letterpaper]{IEEEtran}
\usepackage{amsmath,epsfig}
\usepackage{algorithm}
\usepackage{bm}
\usepackage{tikz}
\usepackage{graphicx}
\usetikzlibrary{positioning}
\usetikzlibrary{shapes.multipart}
\usepackage{pgfplots}
\usepackage[usenames,dvipsnames]{pstricks}
\usepackage{epsfig}
\usepackage{pst-grad} % For gradients
\usepackage{pst-plot} % For axes
\usepackage{subfigure}
\usepackage[capitalize]{cleveref}

\usepackage{subfigure}
\usepackage[T1]{fontenc}
\usepackage{amsmath}
\usepackage{graphics}
\include{psfig}
\usepackage{epsfig}
\usepackage{array}
\usepackage{psfrag}
\usepackage{amssymb}
\usepackage{color}
\usepackage{pstricks,pst-node,pst-text,pst-3d,pst-plot}
\usepackage{cite}

\usepackage{ulem}
\definecolor{dgreen}{rgb}{0, 0.8, 0.1}

\begin{document}

\title{High Resolution Millimeter Wave Imaging Based on FMCW Radar Systems at W-Band}
%\doublespacing

\author{Shahrokh Hamidi$^*$\thanks{Shahrokh Hamidi is with the Faculty of Electrical and Computer
Engineering, University of Waterloo, 200 University Ave W, Waterloo, ON, Canada, N2L 3G1.
e-mail: \texttt{Shahrokh.Hamidi@uwaterloo.ca}.} \and and M.R. Nezhad-Ahmadi$^*$\thanks{M.R. Nezhad-Ahmad is with the Faculty of Electrical and Computer Engineering, University of Waterloo, 200 University Ave W, Waterloo, ON, Canada, N2L 3G1. 
e-mail: \texttt{mrnezhad@uwaterloo.ca}.}}

\maketitle

\thispagestyle{empty}
\pagestyle{plain} 

\begin{tikzpicture}[remember picture, overlay]
      \node[font=\small] at ([yshift=-1cm]current page.north)  {\copyright IEEE};
\end{tikzpicture}

\begin{abstract}
In this paper, we present a unique $\text {2D}$ high resolution, compact, low-cost, lightweight, and highly accurate millimeter wave imagery system capable of operating in all weather conditions. 
We describe millimeter wave imaging process in detail and present several novel signal processing methods with their applications. To create the array, we utilize the Synthetic Aperture Radar (SAR) concept. 
The imagery system presented in this paper, can strongly compete with Lidar systems as the resolution limit is at the same level. Furthermore, in contrast to the Lidar systems, our imagery system can operate in heavy rain and dense
fog and produce high quality images.

Finally, we utilize our wide-band custom-made Frequency Modulated Continuous Wave (FMCW) radar, which operates at W-band with $\text {33 GHz}$ bandwidth, for data collection and present the results. 
\end{abstract}


\section{Introduction}
Radar imaging is an important topic with numerous applications which has been used in different areas over the past decades\cite{Mahafza, Skolnik, Cumming, Soumekh}. The goal in radar imaging is to create 2D or 3D images from the targets located in the radar's field of view. The image is the representation of the complex radar cross section of the targets. In the field of radar imaging, creating high resolution images is of utmost importance. The range resolution is directly related to the bandwidth of the system. However, increasing the bandwidth of the system at low frequencies such as L, C, or X bands, that have been utilized extensively for radar imaging, is  challenging. 
Moreover, enhancing the number of elements to obtain higher resolution in the azimuth direction, enhances the complexity of the system and increases its physical size as well.  

Recently, millimeter, sub-terahertz, and terahertz waves have been used to perform radar imaging \cite{Yanik_2D, Yanik_2D_SAR_MIMO, Yanik_3D_SAR_MIMO, Yanik_3D_SAR_MIMO_, Shahrokh_SAR_1, Shahrokh_SAR_2, Shahrokh_CSAR, 80GHz, 300GHz_1, 300GHz_2,  terahertz_CSAR, sub-terahertz_1, terahertz_2}. Compared to the systems that operate at lower frequencies, the advantage of the systems which operate at millimeter wave, sub-terahertz, and terahertz frequencies is that they are lightweight, compact, and have higher bandwidth. Operating with higher bandwidth will allow the system to obtain higher resolution in the range direction.  In addition, operating at higher frequencies makes it possible to create apertures with large number of elements while the array still occupies a small physical space. This, in turn, yields higher resolution in the azimuth direction.  

The imagery systems operating at  millimeter wave, sub-terahertz, and terahertz frequencies have numerous applications including material characterization, high resolution localization, non-destructive testing (NDT), and near-field imaging, to mention a few.

\subsection{Related Work}
The work presented in \cite{Yanik_2D, Yanik_2D_SAR_MIMO, Yanik_3D_SAR_MIMO, Yanik_3D_SAR_MIMO_,  Shahrokh_SAR_1, Shahrokh_SAR_2, Shahrokh_CSAR} are based on commercially available Frequency Modulated Continuous Wave (FMCW) radar chips that operate at $\text {77 GHz}$ with $\text {4 GHz}$ bandwidth  \cite{Mahafza, Skolnik, Shahrokh_TDM, Shahrokh_CDM}. High resolution and high quality images have been created. However, due to the fact that the bandwidth of the systems is limited to $\text {4 GHz}$, therefore, the best range resolution that we can achieve is approximately $\text {5 cm}$. 

In \cite{80GHz}, the authors have developed a SiGe-Chip-based $\text {80 GHz}$ FMCW radar system with $\text {25.6 GHz}$ bandwidth for high resolution imaging. The work is based on a low power SiGe-MMIC system operating as a FMCW radar. A large  aperture has been created synthetically and high resolution images have been generated.  

Another interesting work reported in the literature  \cite{300GHz_1, 300GHz_2} is an imagery system operating at $\text {300 GHz}$ with more than 40 GHz bandwidth. The system has been used in both  Synthetic Aperture Radar (SAR) and Inverse SAR (ISAR) modes and has been utilized to create high resolution images.  
In \cite{terahertz_CSAR}, a FMCW radar, operating at terahertz frequencies, has been designed to create high resolution images based on Circular SAR (CSAR) technique. 
In \cite{sub-terahertz_1}, a FMCW radar has been designed at $\text {220 GHz}$ with $\text {8 GHz}$ bandwidth. Then, the system has been used to generate high resolution radar images.  

The major issue with imagery systems operating at sub-terahertz and terahertz frequencies is low output power and high free-space path-loss which makes them only suitable for near-field and close-range applications. 

\subsection{Motivation}
In this paper, we present a high resolution imagery system. The system which we present in this paper, is our custom-made wide-band FMCW radar system. We have selected the W-band with frequencies ranging from $\text {75 GHz}$ to $\text {108 GHz}$ with $\text {33 GHz}$ bandwidth. Compared to the systems operating at lower frequencies, the W-band will make it possible to design a compact and lightweight system. In addition, we can reach higher bandwidth which as a result allows to enhance the range resolution limit of the system.  

However, increasing the frequency range beyond the W-band will cause issues such as lower output power and higher free-space path-loss which systems presented in \cite{300GHz_1, 300GHz_2, terahertz_CSAR,  sub-terahertz_1, terahertz_2} suffer from.  
Moreover, the system presented in this paper is a custom-made FMCW radar. FMCW radars are well-known for power efficiency and lower complexity compared to the pulsed-radars. 

In  addition to the system design and image formation, we present several signal processing techniques such as multi-spectral-based imaging, sub-array-based processing, image de-noising, and apodization which distinguishes our work from \cite{Yanik_2D, Yanik_2D_SAR_MIMO, Yanik_3D_SAR_MIMO, Yanik_3D_SAR_MIMO_, Shahrokh_SAR_1, Shahrokh_SAR_2, Shahrokh_CSAR, 80GHz, 300GHz_1,300GHz_2, terahertz_CSAR,  sub-terahertz_1, terahertz_2}. The multi-spectral-based imaging subject paves the way to applications such as material characterization. The sub-array-based processing approach provides approximately $\text {3 dB}$ gain which is a remarkable achievement since the origin of the gain is the signal processing unit.  
We, further, describe the system and the imaging process in detail. Moreover, we present image de-noising. We use the state-of-the-art de-noising techniques from the field of image processing and apply it to the reconstructed images. Since the resolution of the reconstructed images is extremely high, therefore, it is plausible to bring sophisticated methods from the field of image processing in to the field of radar imaging.    
In addition, we utilize apodization technique. The apodization method is performed to reduce the side-lobe levels while preserving the resolution limit of the imagery system \cite{Apodization_1}.

Another major difference between the work we present in this paper and the material reported in \cite{Yanik_2D, Yanik_2D_SAR_MIMO, Yanik_3D_SAR_MIMO, Yanik_3D_SAR_MIMO_, Shahrokh_SAR_1, Shahrokh_SAR_2, Shahrokh_CSAR, 80GHz, 300GHz_1,300GHz_2, terahertz_CSAR,  sub-terahertz_1, terahertz_2} is related to the fact that we have collected data from multiple realistic scenarios in different weather conditions. The main purpose is to show the capability of the imagery system, presented in this work, in producing high resolution images from complex realistic scenarios in different weather conditions such as day, night, rain, and fog. 

\subsection{Organization}
The paper has been organized as follows. In Section \ref{system model}, we present the system model along with the imaging algorithm. Section \ref{Apodization} has been dedicated to the side-lobe reduction using apodization-based methods. In Section \ref{multi-spectral imaging}, we describe the multi-spectral-based imaging technique. Section \ref{image de-noising} discusses the image de-noising process. In Section \ref{sub-array-based processing}, we present the sub-array-based processing method. Finally, section \ref{experimental results} has been dedicated to the experimental results which is followed by the concluding remarks.

\section{System Model}\label{system model}
\begin{figure}
\centering
\includegraphics[height=4cm,width=6cm]{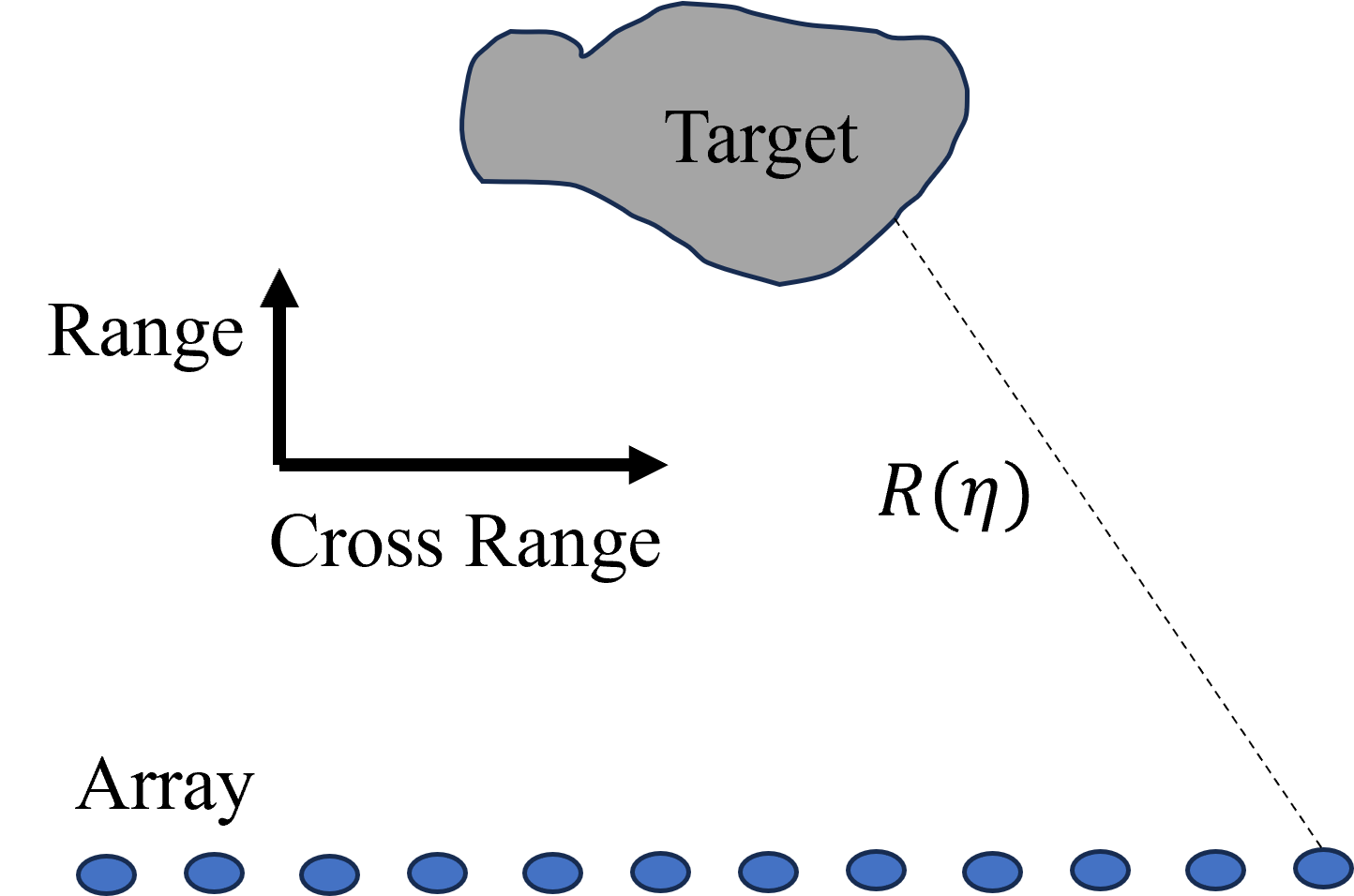}
\caption{The geometry of the model. 
\label{fig:model_geometry}}
\end{figure}

Fig.~\ref{fig:model_geometry} illustrates the geometry of the model that we study in this paper. {The array is created synthetically. The imagery system is a single element TX-RX FMCW radar which is moved along the horizontal direction to create a large array based on the SAR concept.}  
In FMCW radar systems, the signal transmitted by the transmitter and received at the location of the receiver, after being reflected from the $l^{\text {th}}$ point target,  {and at the output of the mixer,} is described as,
\begin{eqnarray}
\label{initial_signal}
s^{(l)}(t, \eta) = \sigma_l \; u \left(\frac{t-\tau_l(\eta)}{T}\right) \; e^{\displaystyle  j 4 \pi (f_c  +  0.5\beta t) R_l(\eta)/c},
\end{eqnarray}
where {$f_c$} is the carrier frequency and  {$\sigma_l$} is the radar cross section for the $l^{\text {th}}$ point target. The parameter {$\beta$} is given as {${b/T}$}, where $b$ and { $T$} stand for the bandwidth and the chirp-time, respectively. Also,  {$u$} represents the envelope of the signal and is considered to be a rectangular window.  Furthermore, {$t$} and {$\eta$} represent the fast- and slow-time parameters, respectively. 
Finally, {$R_{l}$} is the radial distance between the target and the radar.

Upon taking Fourier transform  with respect to the fast-time parameter {$t$} from (\ref{initial_signal}), we obtain 
\begin{eqnarray}
\label{initial_signal_ft}
s^{(l)}(f, \eta) = \sigma_l \; {\text {sinc}\left(T(f-\frac{2R_l(\eta)}{c}\beta)\right) }e^{\displaystyle  j 4 \pi f_c R_l(\eta)/c}.
\end{eqnarray}
This step results in the localization of the energy of the signal in the range direction. The next step is to localize the energy of the signal in the azimuth direction. To accomplish this goal, we implement the back-projection algorithm \cite{Soumekh} as follows
\begin{eqnarray}
\label{BP_algo}
 \hat{\sigma}_l  = \int s^{(l)}(f, \eta)  e^{\displaystyle  - j 4 \pi f_c R_i(\eta)/c} d\eta,
\end{eqnarray}
where $ \hat{\sigma}_l $ is the estimated complex valued radar cross section of the $l^{\text {th}}$ point reflector located at the $i ^{\text {th}}$ pixel.

\section{Apodization}\label{Apodization}
The common technique for side-lobe reduction in radar imaging, is by applying a window to the data in both the range and azimuth directions. However, the side-lobe reduction based on windowing method will reduce the resolution of the system. Apodization is a method which is utilized to decrease the side-lobe levels without changing the resolution limit of the system. In fact, the resolution of the system remains intact while the side-lobe levels are reduced and will be at the same level of the applied window \cite{Apodization_1}.  

Complex dual apodization is yet a more advanced apodization technique which operates on real and imaginary parts of the signal separately and generates better results compared to the apodization method  \cite{Apodization_1}. In this paper, we will be using both the dual apodization and the complex dual apodization techniques to reduce the side-lobe levels while preserving the resolution limit of the imagery system.

\section{Multi-Spectral-Based Imaging}\label{multi-spectral imaging}
In this section, we present multi-spectral-based imaging. The concept is based on partitioning the entire bandwidth into 3 sub-bands. We, then, create 3 separate images and assign red, green, and blue colors to them, respectively. At the end, we display the image in color. 

{In fact, the sub-bands are created using the IF signal (also known as the beat signal). We divide the beat signal, in fast-time domain, into three equal sections where each section represents one-third of the initial signal.  Next, each part is used to create an independent image from the scene. Finally, all three images are assigned red, blue, and green colors, respectively, and the result is plotted.}

\begin{figure}
\centering
%\psfrag{histogram12345678}{\tiny \rm histogram}
\includegraphics[height=6cm,width=8cm]{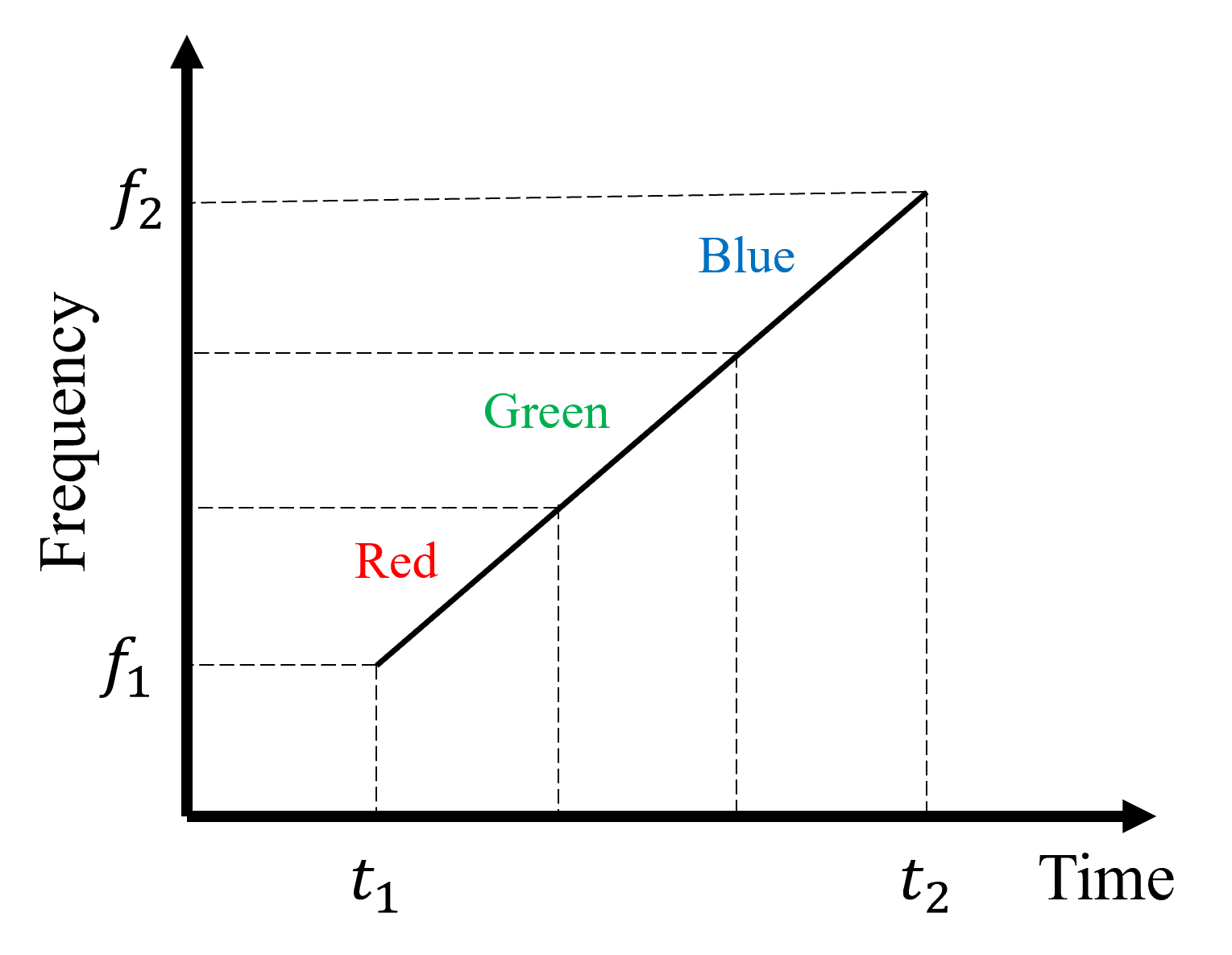}
\caption{Dividing the whole bandwidth of the system into 3 equal sub-bands for multi-spectral-based imaging. 
\label{fig:multi_spectral_diagram}}
\end{figure}

Fig.~\ref{fig:multi_spectral_diagram} shows the process of dividing the entire bandwidth of the system into 3 sub-bands. One of the applications of multi-spectral-based imaging is for materiel characterization in which targets that display the same response over the entire frequency band are presented as white and targets that have frequency-dependent response will be seen as non-white in the processed image.

\section{Image De-noising}\label{image de-noising}
The image created by the system is contaminated by the noise of the system. In this section, we address the image de-noising process. The goal of image de-noising is to remove the noise while the edges as well as the fine structures in the image are preserved. Furthermore, image smearing and blurring will not occur. Image de-noising assists with observing the fine details of the image which have been covered by the noise of the system. Since the imagery system, presented in this paper, creates incredibly high resolution images, which is comparable to optical images in resolution and quality, therefore, we will be using image de-noising techniques from the filed of image processing. 

To perform the image de-noising process, we implement the $\rm l1$-TV method \cite{TV_Rudin} which is based on the following convex optimization problem,  
\begin{equation}
\begin{aligned}
\label{denoising_opt}
\min_{u} \quad & \int_{\Omega} ||\nabla u||_1\, dx + \frac{\lambda}{2} \; \int_{\Omega} (f - u)^2\, dx.
\end{aligned}
\end{equation}
In (\ref{denoising_opt}), {$f$} represents the noisy image and $\lambda$ is the hyper-parameter of the optimization problem. Furthermore, $\Omega$ represents the image domain. 

The optimization problem given in  (\ref{denoising_opt}) is convex \cite{Convex_Boyd} and can be solved using existing packages such as $\text {cvx}$ in MATLAB or $\text {cvxpy}$ in Python. However, for images with large number of pixels, solving the optimization problem, given in (\ref{denoising_opt}), is computationally prohibited.

To mitigate this issue, we will be solving the Euler-Lagrange equation \cite{math_book} for the optimization problem given in (\ref{denoising_opt}). Moreover, to make the algorithm resilient to edge smearing, we will utilize an anisotropic version of (\ref{denoising_opt}).
Consequently, upon using the  Euler-Lagrange equation we obtain the following nonlinear elliptic partial differential equation,

\begin{equation}
\begin{cases}
\label{denoising_opt_EL}
& \displaystyle \frac{\partial u}{\partial t}  = \nabla . \left({c(|\nabla (G_\sigma * u)|^2)}{\nabla u}\right ), \\ 
& u(x,0) = f (x), \\
& c(||\nabla I||) = e^{-\displaystyle {\left(||\nabla I||/K\right)}^2}, \\
& G_\sigma(I) = \sigma^{-1/2} e^{\displaystyle -|I|^2/(4 \sigma)}, \\
& (t,x) \in (0,T) \times \Omega.
\end{cases}
\end{equation}
In (\ref{denoising_opt_EL}), the parameter K controls the sensitivity to the edges and is a hyper-parameter and $(*)$ represents the convolution operator. 
The algorithm presented in (\ref{denoising_opt_EL}) is, in fact, the nonlinear diffusion method, also known as modified Perona-Malik diffusion technique \cite{Penora}, which is used for image de-noising.
 
We will implement (\ref{denoising_opt_EL}) directly in Python to mitigate the effect of the noise. One of the main features of the de-noising technique given in (\ref{denoising_opt_EL}) is the ability of the algorithm to preserve the edges and fine details of the image while being able to remove the effect of the noise.

\section{Sub-Array-Based Processing}\label{sub-array-based processing}
\begin{figure}
\centering
%\psfrag{histogram12345678}{\tiny \rm histogram}
\includegraphics[height=6cm,width=6cm]{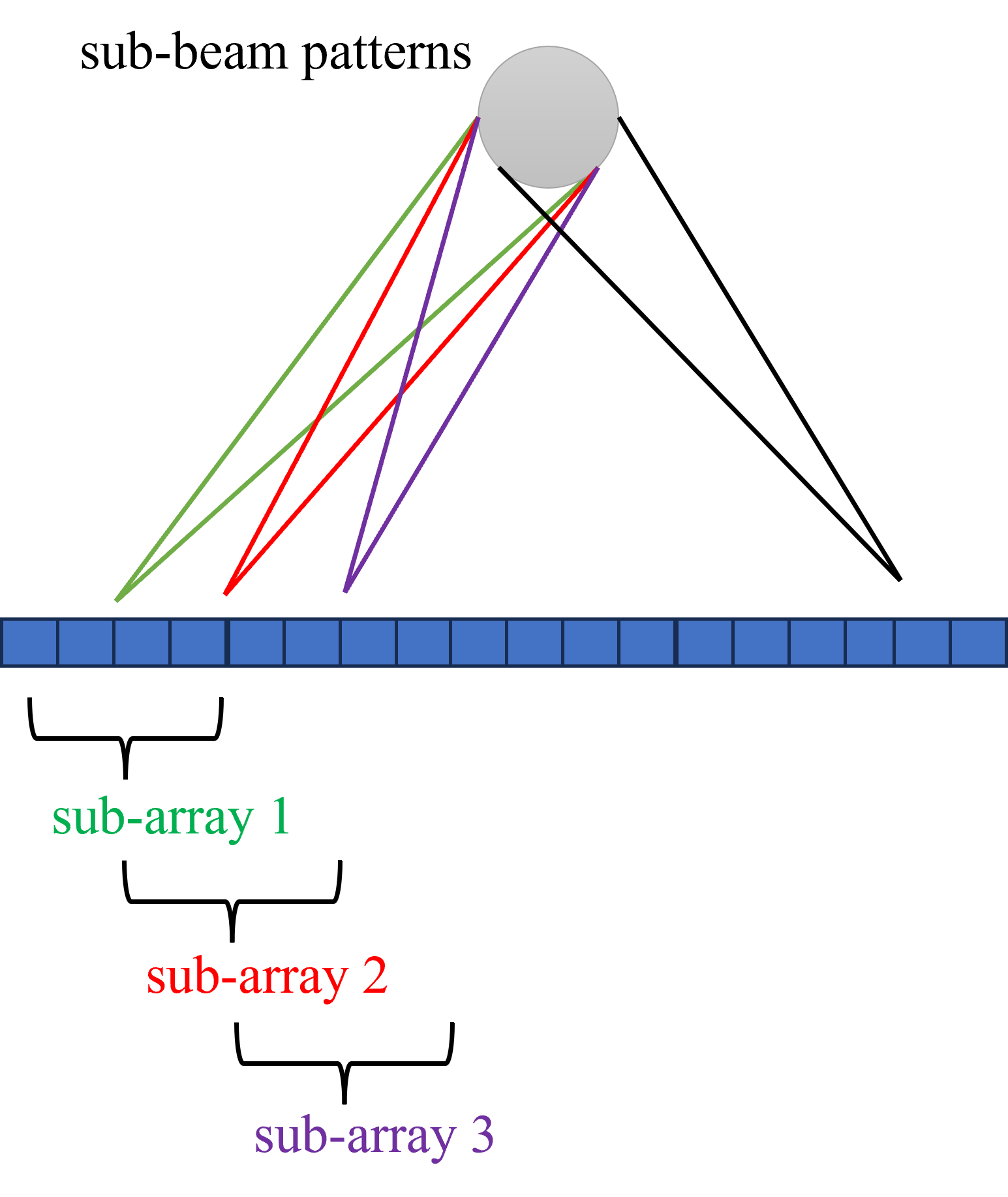}
\caption{Partitioning the entire array into sub-arrays and directing the sub-beams toward a specific place in the field of view of the imagery system which will be used to perform the final stage of the beam-forming process. 
\label{fig:sub_array_processing}}
\end{figure}

In this section, we discuss sub-array-based processing. The goal of the sub-array-based method is to increase the Signal to Noise Ratio (SNR) which is a crucial factor for image reconstruction at millimeter wave frequencies. To begin with, we partition the array into smaller sub-arrays. Then, we perform beam-forming per each sub-array and direct the sub-beams toward a specific spot in the field of view of the imagery system. Next, we utilize all the sub-beams and perform the final beam-forming stage over all the sub-beams that we have created and generate the final image. Fig.~\ref{fig:sub_array_processing} illustrates the sub-array-based imaging concept.

The process of creating the sub-arrays will increase the SNR which, in turn, will reduce the noise floor which is, indeed, the main goal for the implementation of the sub-array-based approach. 
Nevertheless, since the sub-arrays are created with non-zero overlap, which forces the samples of the noise to become correlated, therefore, the gain that we achieve by performing the sub-array-based processing, is not exactly equal to the number of the elements in the sub-arrays. However, we can still obtain approximately 3 dB gain in SNR which is considered to be a remarkable achievement.  It should be noted that, the gain is purely obtained from the signal processing unit and that is what makes this approach valuable.  In other words, this approach does not increase the complexity of the imagery system. 

{It is worth mentioning that, increasing the number of elements in the sub-array results in narrower field of view for the imagery system.} 

\section{Experimental Results}\label{experimental results}
In this section, we present the experimental results. We have designed a FMCW radar at W-band which operates at frequencies ranging from $\text {75 GHz}$ to $\text {108 GHz}$ with $\text {33 GHz}$ bandwidth.
\begin{figure*}
\centering
\begin{tikzpicture}
 \node (img1)  {\includegraphics[height=6cm,width=14cm]{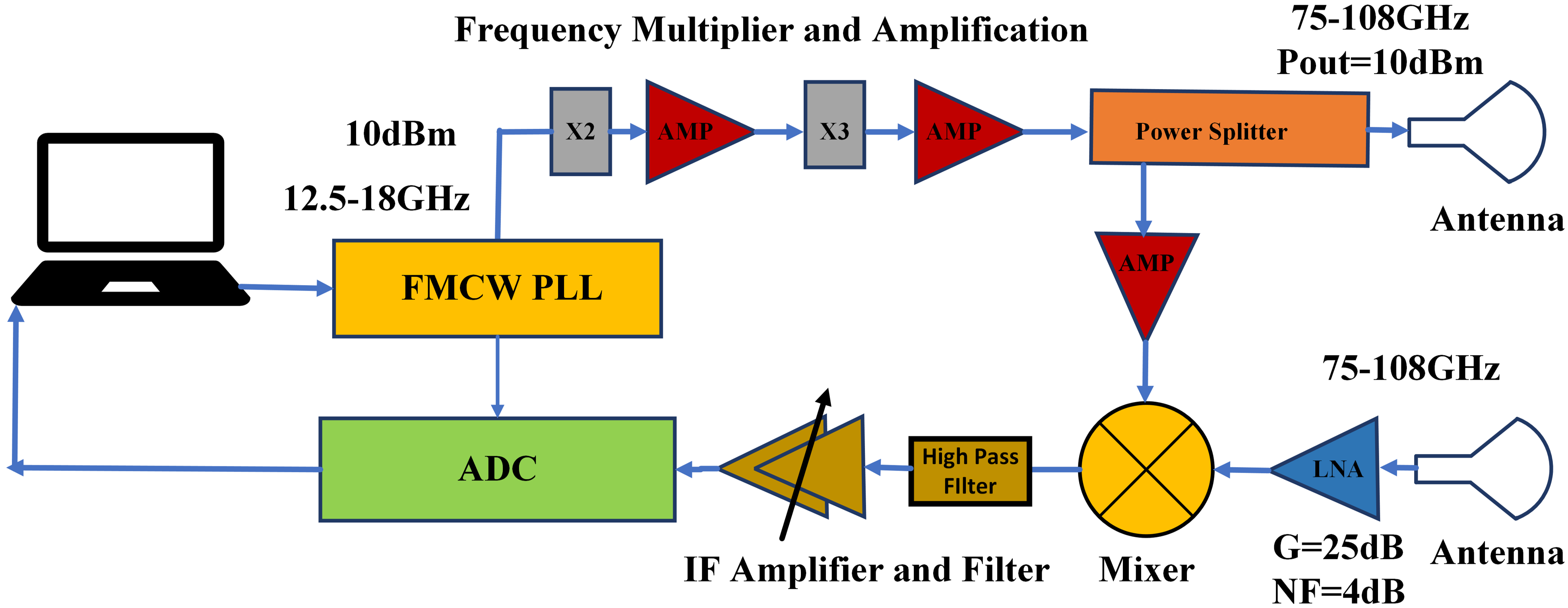}};
\end{tikzpicture}
\caption{The block diagram of the FMCW radar system.}
 \label{fig:block_diagram}
\end{figure*}
Fig.~\ref{fig:block_diagram} shows the block diagram of the developed  broadband imaging radar system incorporating FMCW technology, based on a Phase-Locked Loop (PLL) FMCW generator. This generator outputs a signal spanning a broad initial frequency spectrum of $\text {12.5 GHz}$ to $\text {18 GHz}$, with a power specification of $\text {10 dBm}$. Subsequent frequency multiplication through a configured doubler and tripler extends the signal's frequency reach, achieving an expansive band from $\text {75 GHz}$ to $\text {108 GHz}$.

This frequency multiplication process enhances the bandwidth of the radar system which is critical for achieving high-resolution imaging. Amplification stages subsequent to each multiplication step ensure proper signal power at the output and maintain the broadband characteristic of the output.
A power splitter, serving both the transmission arm and the local oscillator input for the mixer in the receive arm. The transmission arm leverages the broadband capabilities to radiate a signal through either an open wave-guide or a horn antenna depending on the scenario under test to support high-resolution imaging applications.

The receiving arm utilizes a Low Noise Amplifier (LNA) with  $\text {25 dB}$ gain and  $\text {4 dB}$ noise figure to amplify the reflected broadband signals. These signals, upon mixing with the local oscillator signal, yield an Intermediate Frequency (IF) that contains the beat frequencies from the different targets.

To assure the integrity of the IF signal, a high-pass filter is employed to discard undesirable low-frequency components, with subsequent stages of amplification and filtering enhancing the signal quality. The IF signal is then converted to digital format via an Analog-to-Digital Converter (ADC), making it amenable to perform complex digital processing algorithms.
{The purpose of the high pass filter is to remove or decrease the energy of the close-by targets that are within the minimum range of the radar. Furthermore, the high pass filter blocks the leakage from the TX into RX and this way it protects the receiver from saturation and it also increases the dynamic range of the system.} 
{The ADC block contains the ADC as well as the low pass Anti Aliasing Filter (AAF).} 
{The sampling frequency for the ADC is $\text {5 MHz}$. The chirp time has been set to $\text {5 msec}$}. 
\begin{figure}
\centering
\begin{tikzpicture}[yshift=0.00001cm][font=\Large]
  \node (img1)  {\includegraphics[height=5cm,width=7cm]{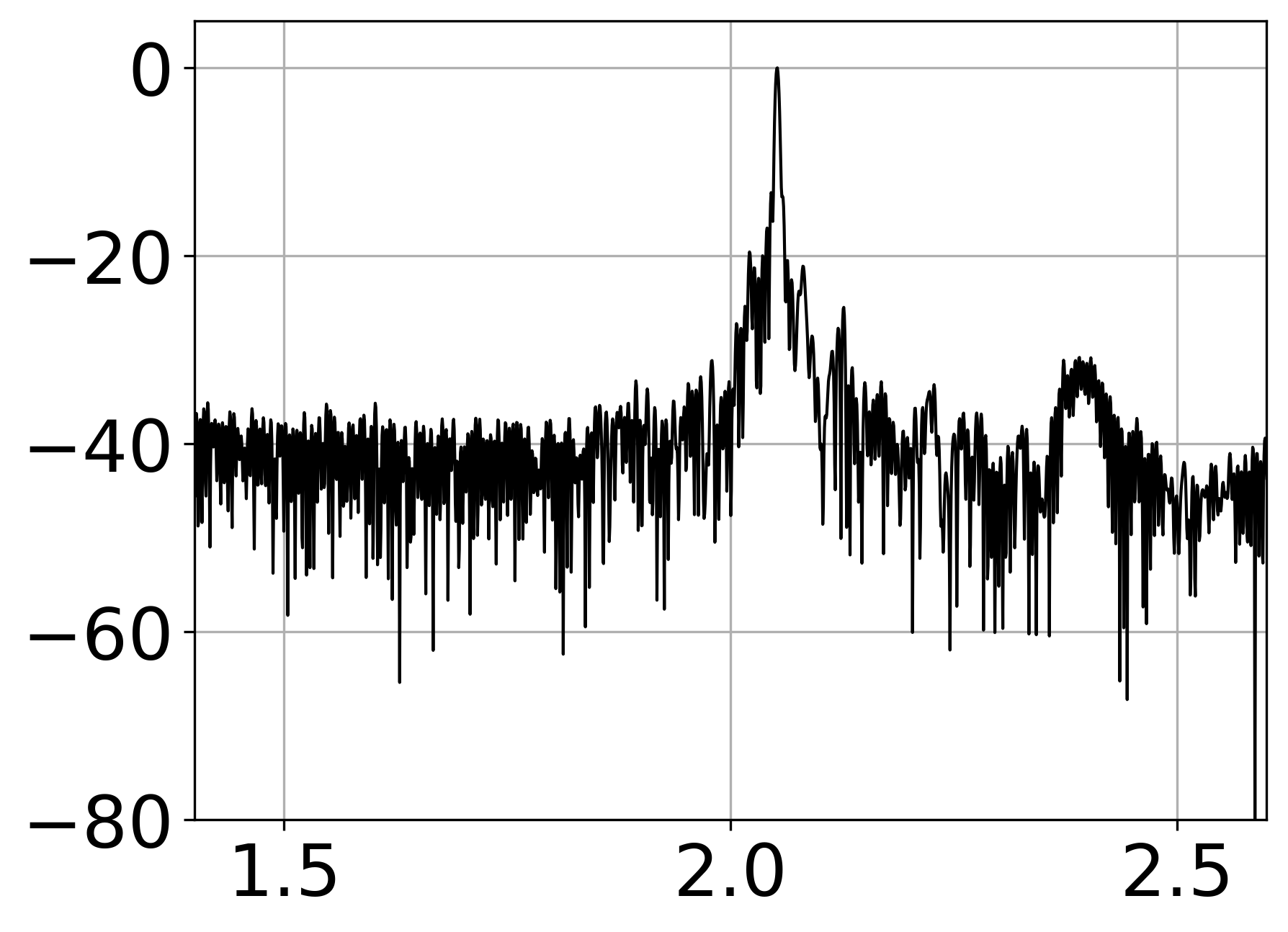}};
  \node[left=of img1, node distance=0cm, rotate = 90, xshift=2.5cm, yshift=-0.8cm,font=\color{black}] {{Normalized Power [dB]}};
  \node[below=of img1, node distance=0cm, xshift=0cm, yshift=1.3cm,font=\color{black}] {{Range [m]}};
 \node[below=of img1, node distance=0cm, xshift=0cm, yshift=0.7cm,font=\color{black}] {{(a)}};
\end{tikzpicture}

\begin{tikzpicture}[yshift = 0.05cm][font=\Large][x=1.0cm,y=1.0cm,scale=1][baseline={(0,0)}]
  \node (img1)  {\includegraphics[height=5cm,width=7cm]{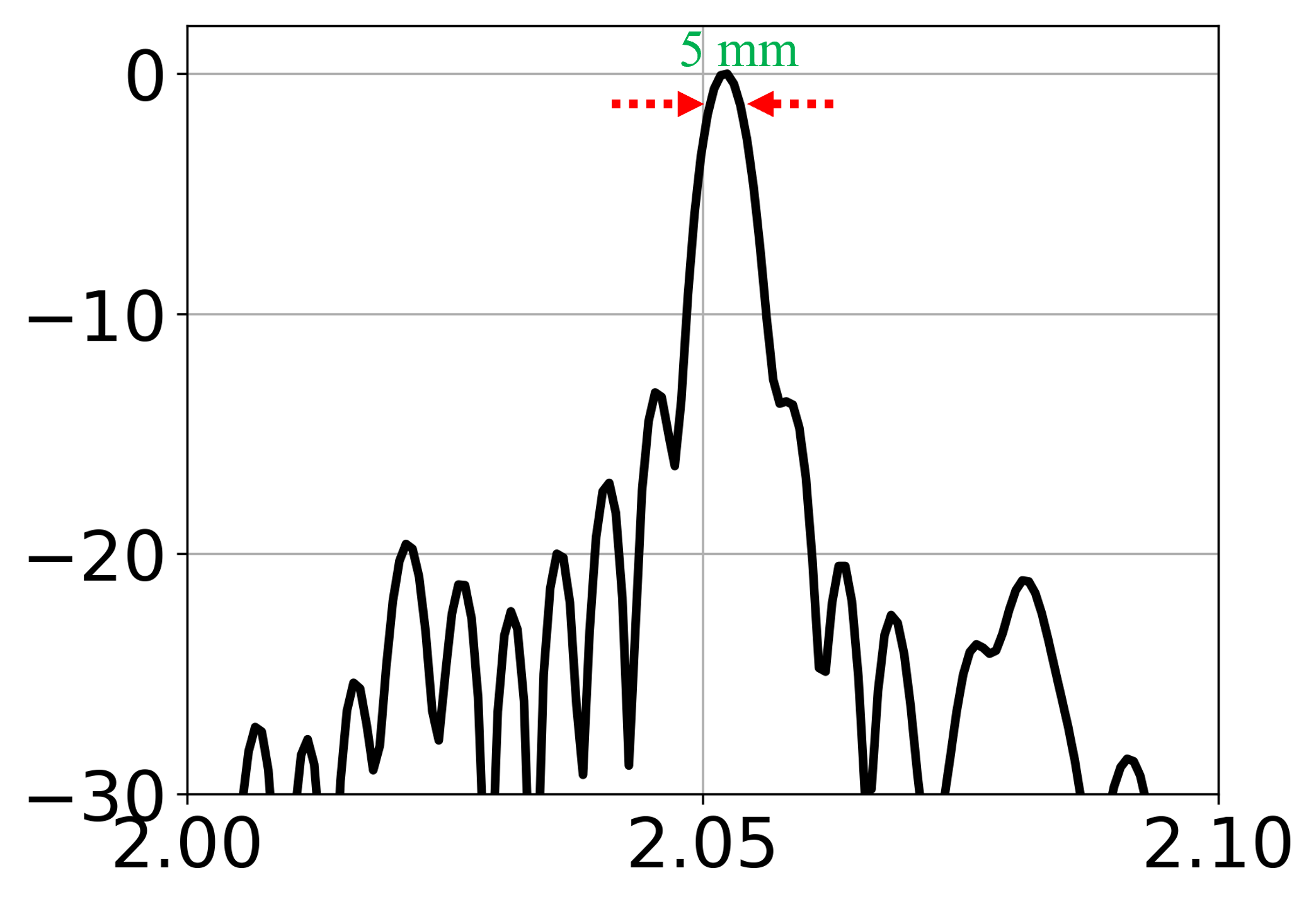}};
  \node[left=of img1, node distance=0cm, rotate = 90, xshift=2.5cm, yshift=-0.8cm,font=\color{black}] {{Normalized  Power [dB]}};
  \node[below=of img1, node distance=0cm, xshift=0cm, yshift=1.3cm,font=\color{black}] {{Range [m]}};
 \node[below=of img1, node distance=0cm, xshift=0cm, yshift=0.7cm,font=\color{black}] {{(b)}};
\end{tikzpicture}
\caption{a) the impulse response of the system using a corner reflector at $\text {2.05 m}$ radial distance, b) the 3 dB beam-width limit. \label{fig:trihedral}}
\label{fig:trihedral}
\end{figure}

In Fig.~\ref{fig:trihedral}, we have shown the impulse response of the system using a corner reflector located at $\text {2.05 m}$ radial distance. 
To perform the image reconstruction process, we have selected the frequency components that rang from $\text {78 GHz}$ to $\text {102 GHz}$ which is the flat part of the spectrum that the system provides and is equivalent to $\text {24 GHz}$ bandwidth. The $\text {24 GHz}$ bandwidth will provide us with $\text {6.3 mm}$ resolution in the range direction. The length of the array, for all the experiments that we present in this section, is $\text {40 cm}$. 

{Fig.~\ref{fig:power_spectrum} illustrates the power spectrum of the system over the entire bandwidth ranging from $\text {75 GHz}$ to $\text {108 GHz}$. From Fig.~\ref{fig:power_spectrum}, it is clear that the system is experiencing large power fluctuation in the beginning and at the end of the spectrum. For the imaging process, we have selected the part of the spectrum corresponding to the range of frequencies from $\text {78 GHz}$ to $\text {102 GHz}$ which is equivalent to $\text {24 GHz}$ bandwidth. 
\begin{figure}
\centering
\begin{tikzpicture}[yshift=0.00001cm][font=\Large]
  \node (img1)  {\includegraphics[height=5cm,width=8cm]{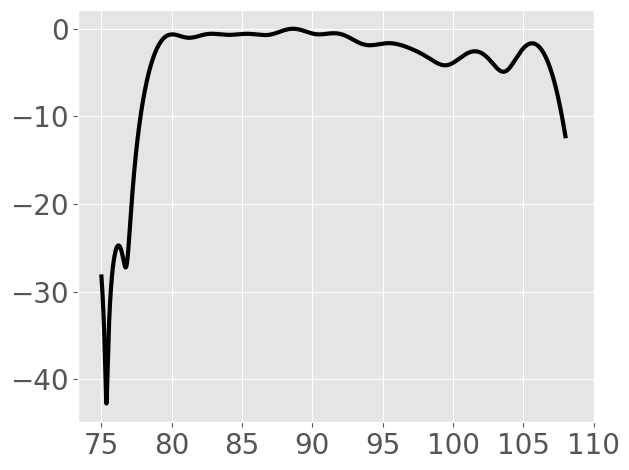}};
  \node[left=of img1, node distance=0cm, rotate = 90, xshift=2.5cm, yshift=-0.9cm,font=\color{black}] {{Normalized Power [dB]}};
  \node[below=of img1, node distance=0cm, xshift=0cm, yshift=1.3cm,font=\color{black}] {{Frequency [GHz]}};
\end{tikzpicture}
\caption{The power spectrum for the system over the entire bandwidth ranging from $\text {75 GHz}$ to $\text {108 GHz}$.}
\label{fig:power_spectrum}
\end{figure}
}
\begin{figure}
\centering
\begin{tikzpicture}[yshift=0.00001cm][font=\Large]
  \node (img1)  {\includegraphics[height=5cm,width=6cm]{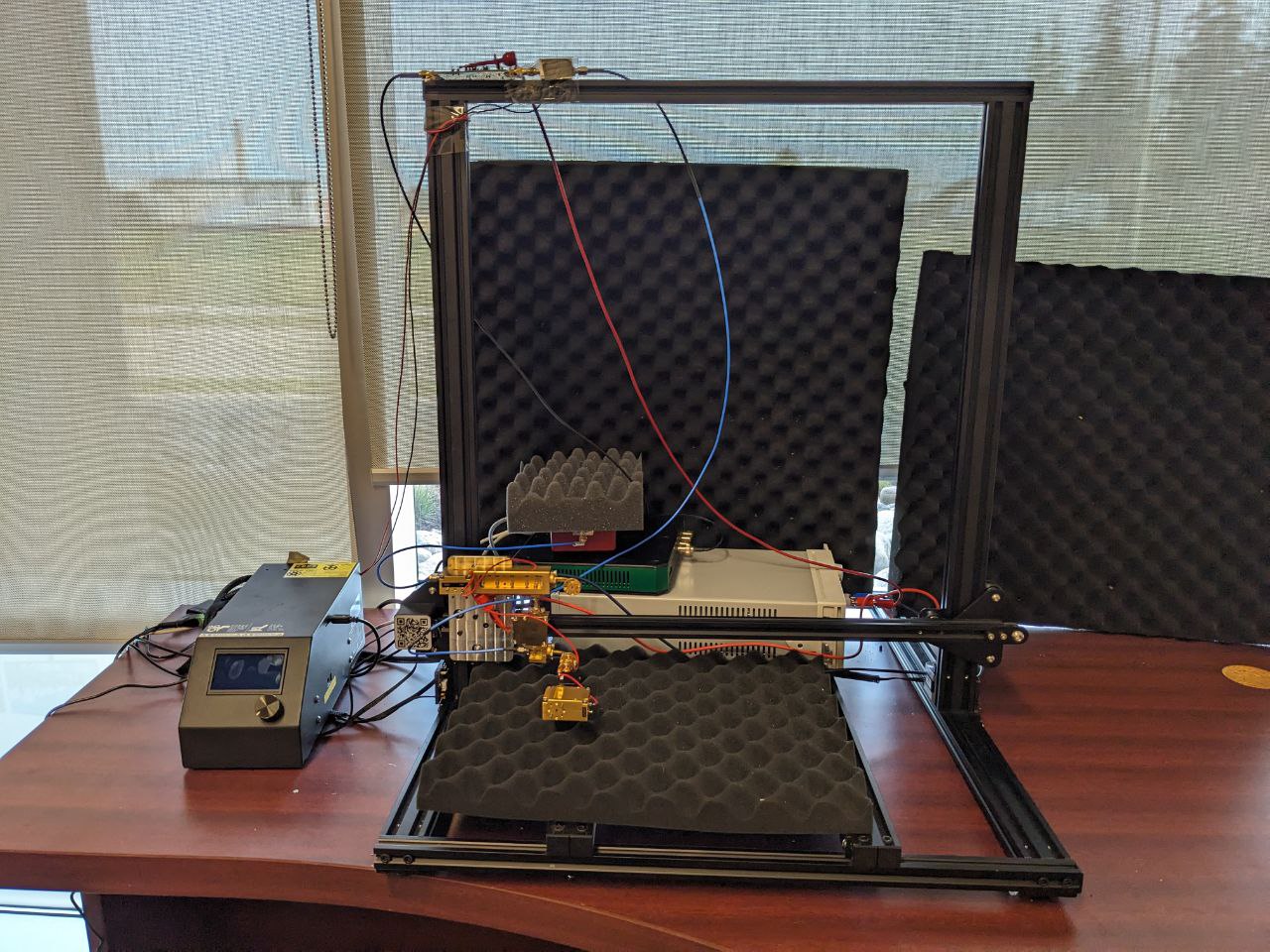}};
\end{tikzpicture}
\caption{The radar system along with the motorized scanner to create the synthetic aperture.}
\label{fig:Radar_system}
\end{figure}
Fig.~\ref{fig:Radar_system} illustrates the radar system  along with the motorized scanner to create the synthetic aperture. 
\begin{figure}
\centering
\begin{tikzpicture}[yshift=0.05cm][font=\Large]
  \node (img1)  {\includegraphics[height=5cm,width=6cm]{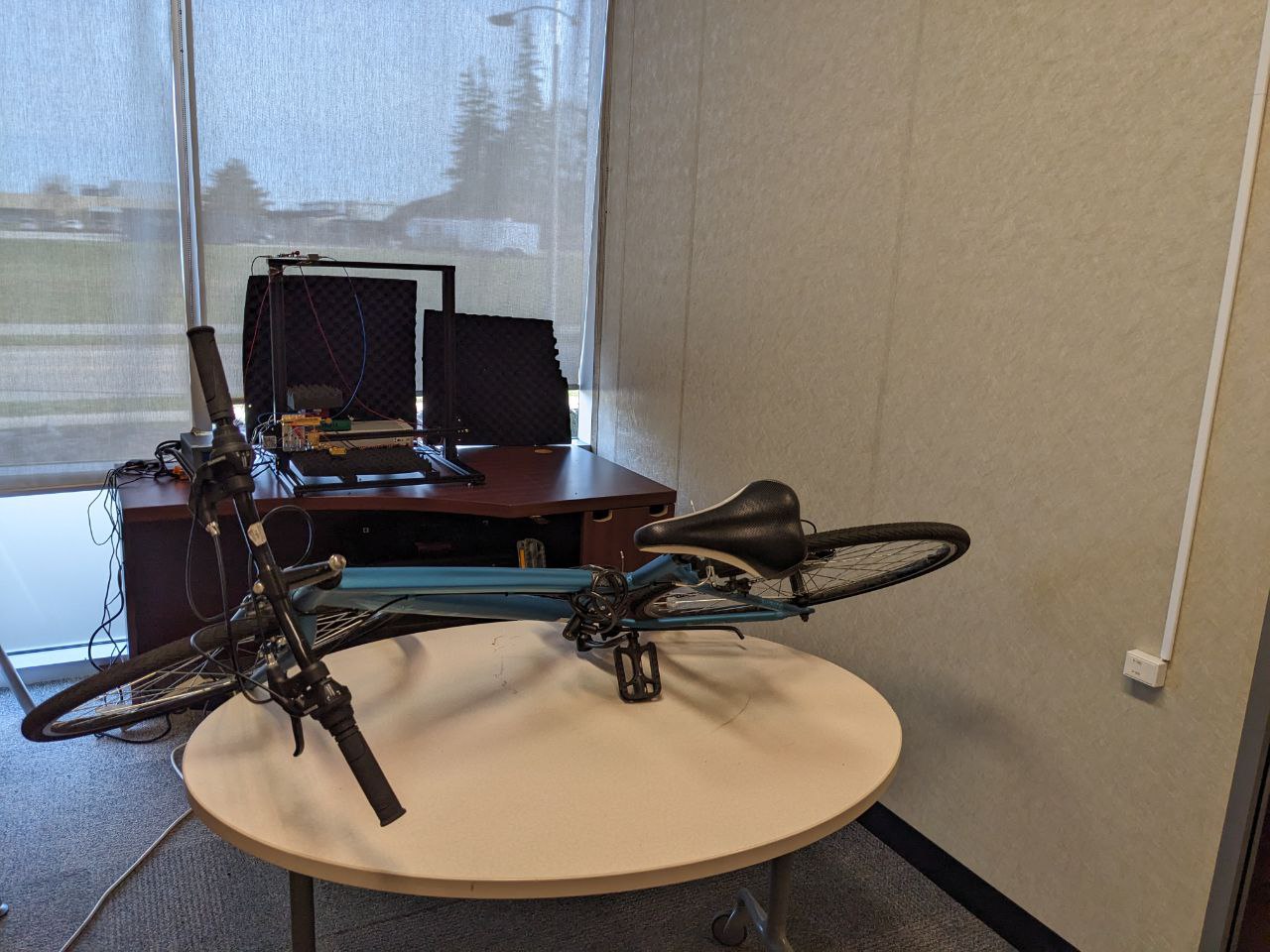}};
\end{tikzpicture}

\caption{The experimental set-up.}
\label{fig:set_up}
\end{figure}
Fig.~\ref{fig:set_up} depicts the experimental set-up in which a bike has been located in front of the radar system.

\subsection{Image Formation}
In this subsection, we present the result of applying the back-projection algorithm to the data-set collected from a bike in front of the imagery system. 
\begin{figure}
\centering
\begin{tikzpicture}[yshift=0.00001cm][font=\Large]
  \node (img1)  {\includegraphics[height=6cm,width=8cm]{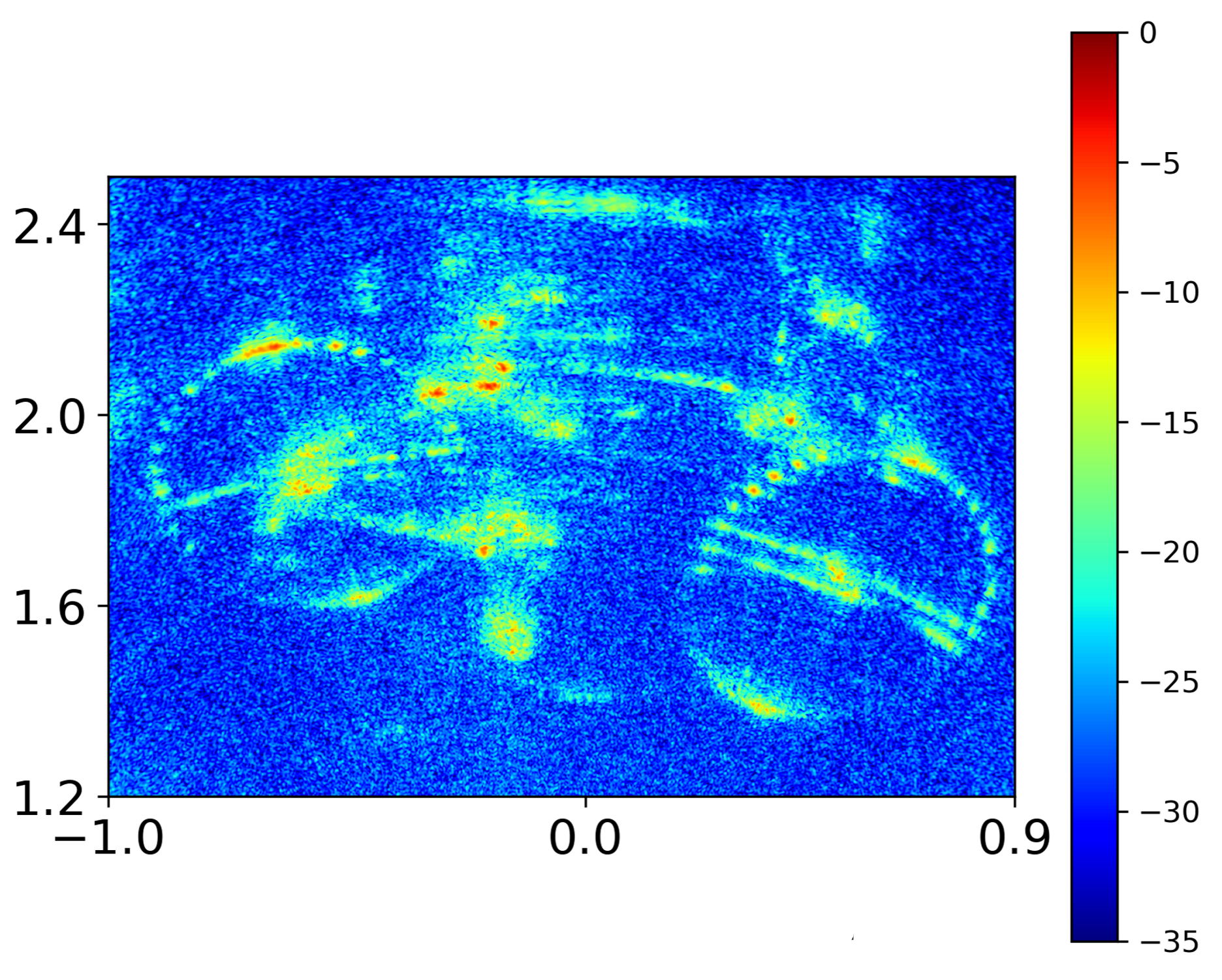}};
   \node[left=of img1, node distance=0cm, rotate = 90, xshift=1cm, yshift=-0.9cm,font=\color{black}] {{Range [m]}};
  \node[below=of img1, node distance=0cm, xshift=0cm, yshift=1.9cm,font=\color{black}] {{Cross Range [m]}};
\end{tikzpicture}

\caption{The result of applying the back-projection algorithm, based on (\ref{BP_algo}), to the experimental data gathered from the bike.}
\label{fig:image_formation}
\end{figure}
Fig.~\ref{fig:image_formation} shows the result of applying the back-projection algorithm, based on (\ref{BP_algo}), to the experimental data gathered from the bike.
\begin{figure}
\centering
\begin{tikzpicture}[yshift=0.05cm][font=\Large]
  \node (img1)  {\includegraphics[height=6cm,width=8cm]{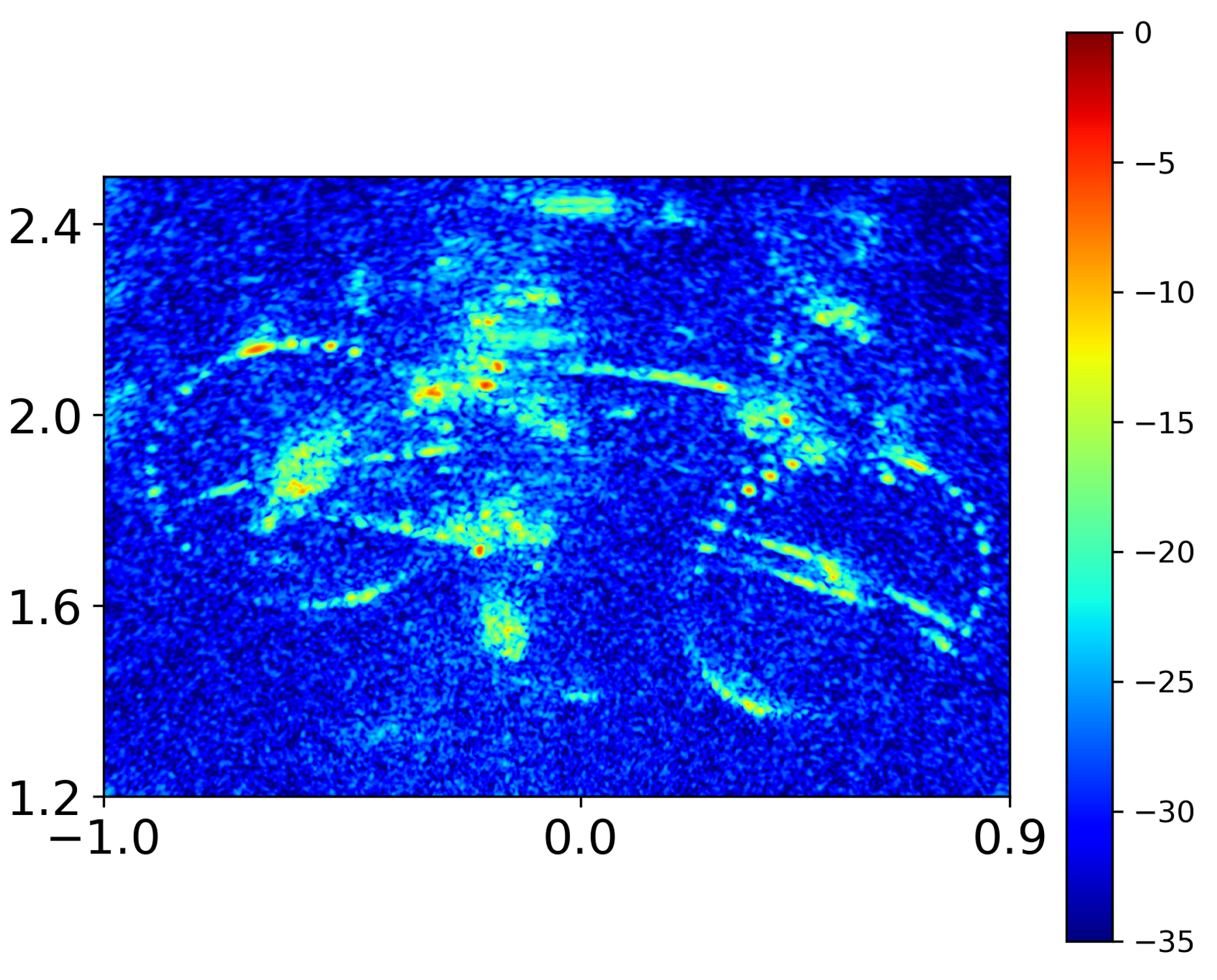}};
 \node[left=of img1, node distance=0cm, rotate = 90, xshift=1cm, yshift=-0.9cm,font=\color{black}] {{Range [m]}};
  \node[below=of img1, node distance=0cm, xshift=0cm, yshift=1.9cm,font=\color{black}] {{Cross Range [m]}};
\end{tikzpicture}
\caption{The reconstructed image after applying Hamming window in both the range and cross-range directions.}
\label{fig:image_formation_window}
\end{figure}
Fig.~\ref{fig:image_formation_window} illustrates the reconstructed image while we have applied Hamming window in both range and cross-range directions to reduce the side-lobe levels. From Fig.~\ref{fig:image_formation} and Fig.~\ref{fig:image_formation_window}, we can see all the details of the bike such as the saddle, tires, pedal, chain, spokes, gears, handlebars, and crossbar.   
\begin{figure}
\centering
\begin{tikzpicture}[yshift=0.00001cm][font=\Large]
  \node (img1)  {\includegraphics[height=6cm,width=6cm]{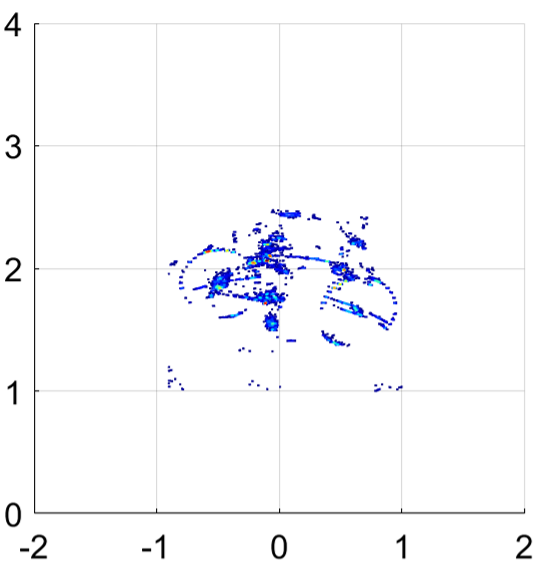}};
   \node[left=of img1, node distance=0cm, rotate = 90, xshift=1cm, yshift=-0.9cm,font=\color{black}] {{Range [m]}};
  \node[below=of img1, node distance=0cm, xshift=0cm, yshift=1.3cm,font=\color{black}] {{Cross Range [m]}};
\end{tikzpicture}

\caption{The point-cloud for the image shown in Fig.~\ref{fig:image_formation}, the information about the intensity is included.}
\label{fig:image_point_cloud_1}
\end{figure}
In Fig.~\ref{fig:image_point_cloud_1}, we have shown the point-cloud for the reconstructed image shown in Fig.~\ref{fig:image_formation}. 

Fig.~\ref{fig:image_point_cloud_1} includes the information  about the intensity of the point reflectors.
\begin{figure}
\centering
\begin{tikzpicture}[yshift=0.05cm][font=\Large]
  \node (img1)  {\includegraphics[height=6cm,width=6cm]{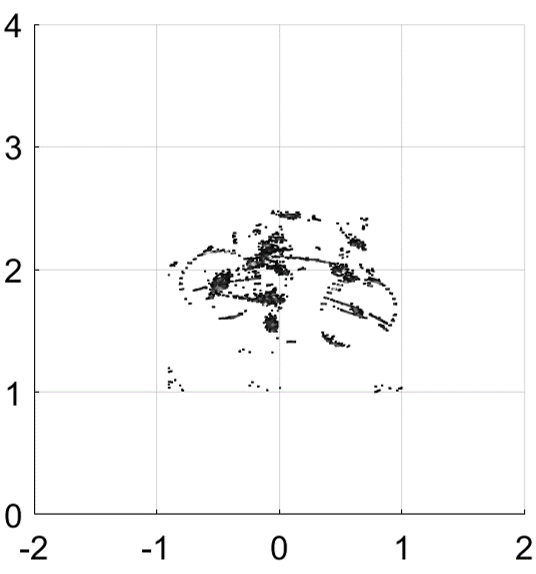}};
 \node[left=of img1, node distance=0cm, rotate = 90, xshift=1cm, yshift=-0.9cm,font=\color{black}] {{Range [m]}};
  \node[below=of img1, node distance=0cm, xshift=0cm, yshift=1.3cm,font=\color{black}] {{Cross Range [m]}};
\end{tikzpicture}

\caption{The point-cloud for the image shown in Fig.~\ref{fig:image_formation} with all points represented at the same intensity level.}
\label{fig:image_point_cloud_2}
\end{figure}
In Fig.~\ref{fig:image_point_cloud_2}, we have shown all the point reflectors with the same intensity level.

\subsection{Apodization}
In this sub-section, the result of the dual apodization and complex dual apodization methods are presented. We have used rectangular and Hamming window to perform the apodization method. The rectangular window represents the image without windowing. It is important to mention that multiple windows can be used for apodization which create multi-layer algorithm for side-lobe reduction while the resolution limit of the system is preserved.

The dual apodization method is based on the absolute value of the reconstructed image and provides a reasonable result. However, the complex dual apodization operates on the real and imaginary parts of the reconstructed image separately and yields an incredibly better result with significantly lower side-lob levels.

\begin{figure}
\centering
\begin{tikzpicture}[yshift=0.00001cm][font=\Large]
  \node (img1)  {\includegraphics[height=6cm,width=8cm]{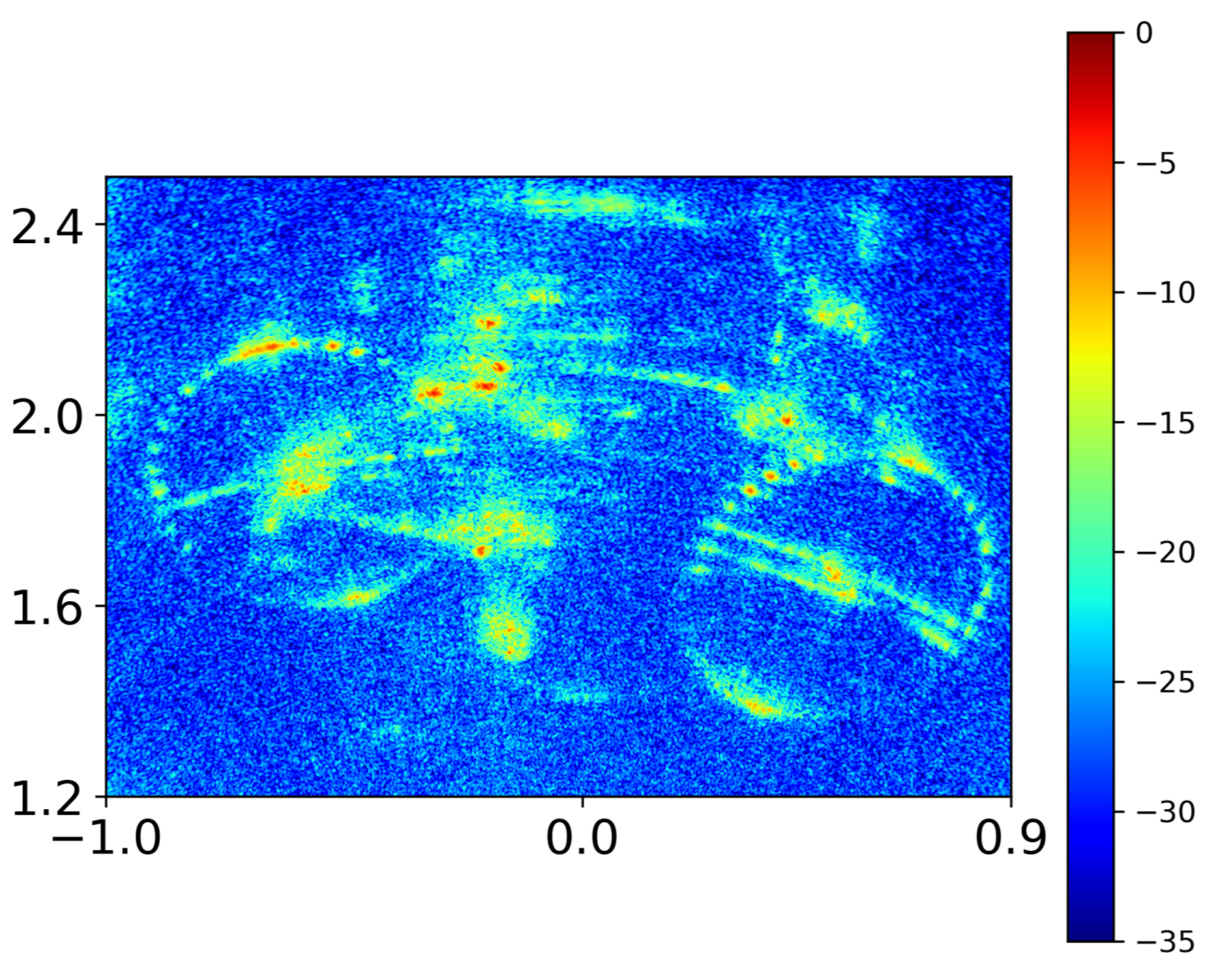}};
   \node[left=of img1, node distance=0cm, rotate = 90, xshift=1cm, yshift=-0.9cm,font=\color{black}] {{Range [m]}};
  \node[below=of img1, node distance=0cm, xshift=0cm, yshift=1.9cm,font=\color{black}] {{Cross Range [m]}};
 \node[below=of img1, node distance=0cm, xshift=0cm, yshift=1.2cm,font=\color{black}] {{(a)}};
\end{tikzpicture}

\begin{tikzpicture}[yshift=0.05cm][font=\Large]
  \node (img1)  {\includegraphics[height=6cm,width=8cm]{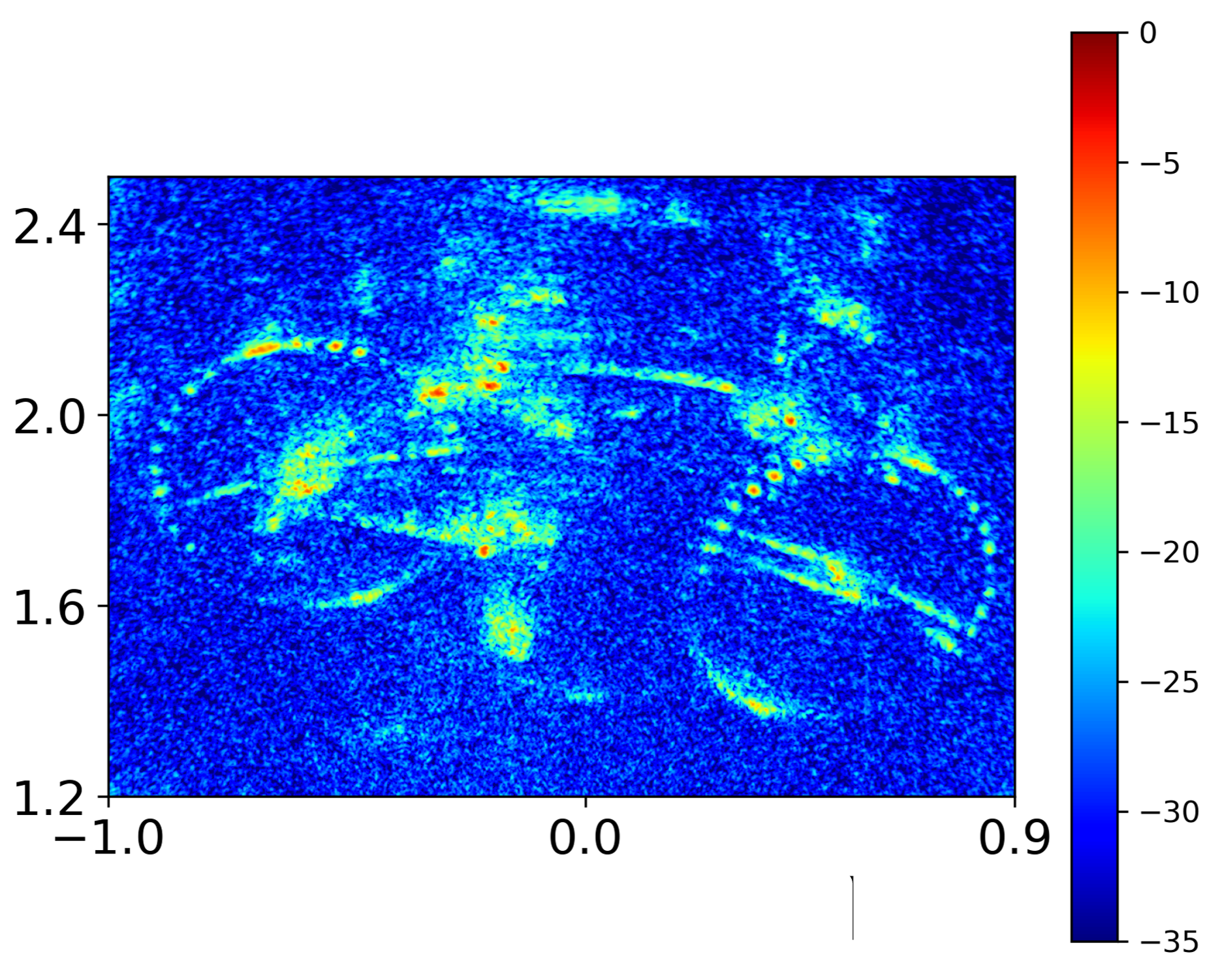}};
 \node[left=of img1, node distance=0cm, rotate = 90, xshift=1cm, yshift=-0.9cm,font=\color{black}] {{Range [m]}};
  \node[below=of img1, node distance=0cm, xshift=0cm, yshift=1.9cm,font=\color{black}] {{Cross Range [m]}};
\node[below=of img1, node distance=0cm, xshift=0cm, yshift=1.2cm,font=\color{black}] {{(b)}};
\end{tikzpicture}

\begin{tikzpicture}[yshift=0.05cm][font=\Large]
  \node (img1)  {\includegraphics[height=6cm,width=8cm]{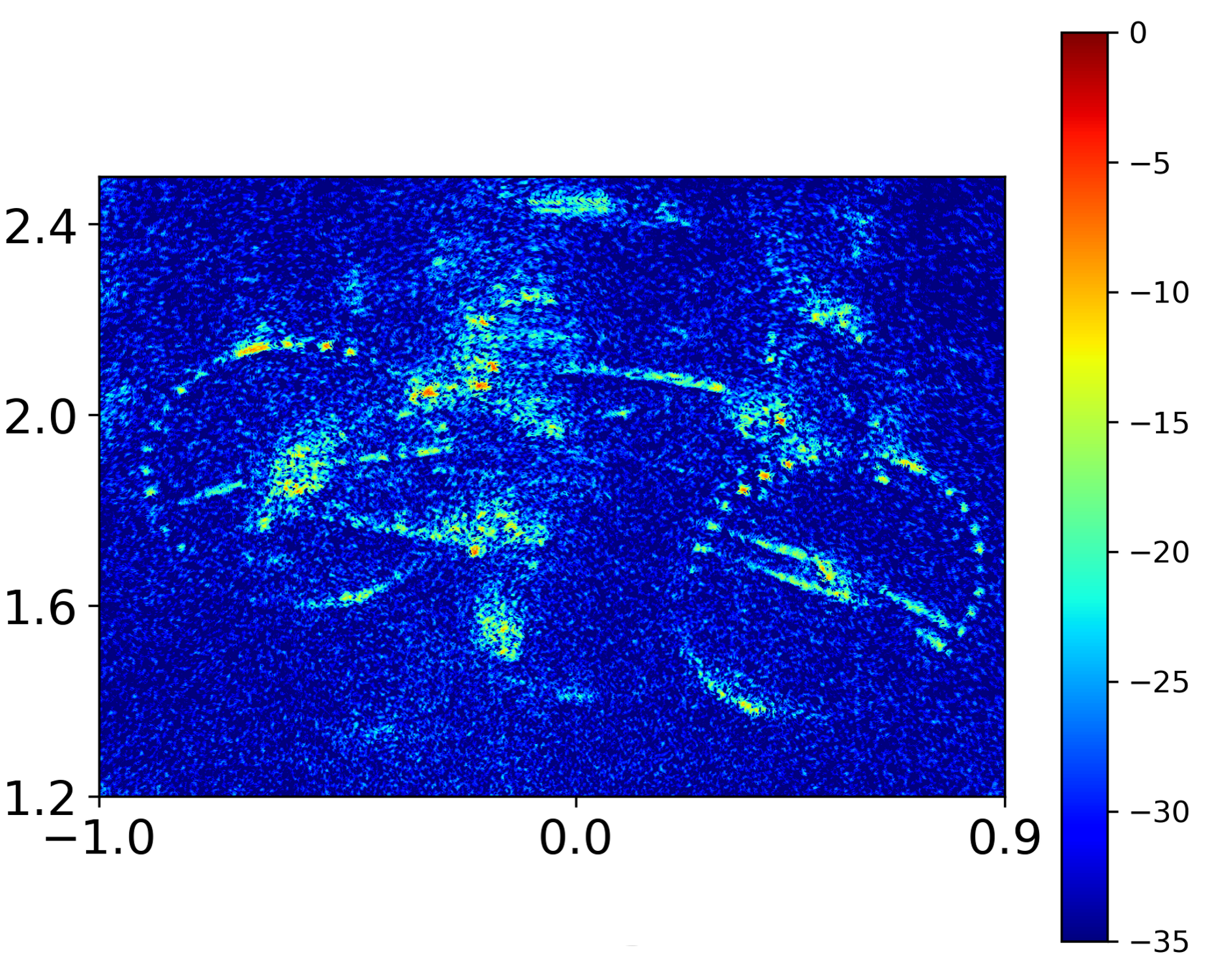}};
 \node[left=of img1, node distance=0cm, rotate = 90, xshift=1cm, yshift=-0.9cm,font=\color{black}] {{Range [m]}};
  \node[below=of img1, node distance=0cm, xshift=0cm, yshift=1.9cm,font=\color{black}] {{Cross Range [m]}};
\node[below=of img1, node distance=0cm, xshift=0cm, yshift=1.2cm,font=\color{black}] {{(c)}};
\end{tikzpicture}

\caption{a) the resulting image without apodization, b) the reconstructed image after applying the dual apodization method, c)  the reconstructed image after applying the complex dual apodization method.}
\label{fig:image_apodization}
\end{figure}
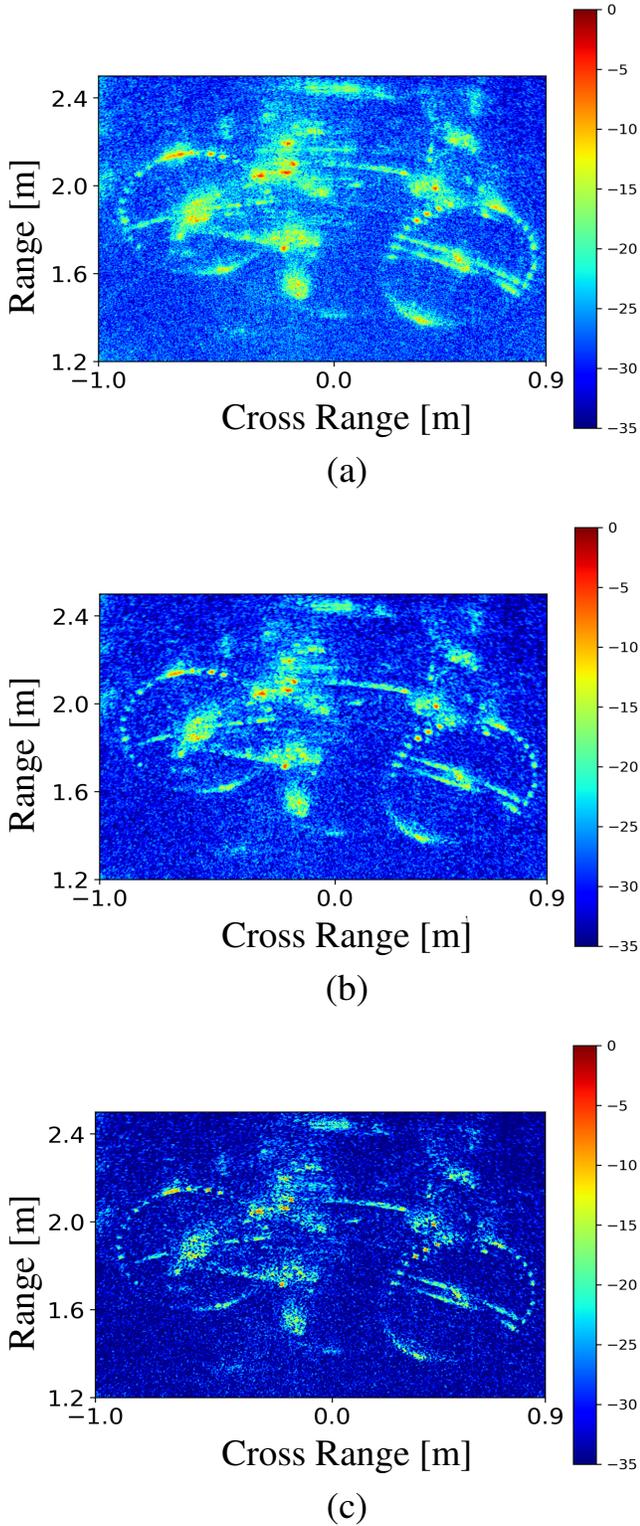

Fig.~\ref{fig:image_apodization} shows the result of utilizing the apodization methods.  Fig.~\ref{fig:image_apodization}-(a) illustrates the image without apodization. In Fig.~\ref{fig:image_apodization}-(b), the result of dual apodization which has been applied to the image is shown. As can be seen from  Fig.~\ref{fig:image_apodization}-(b), the side-lobe levels are lower and the resolution limit is intact. The side-lobe levels will be at the same level as the case in which the Hamming window is used.   

Finally, Fig.~\ref{fig:image_apodization}-(c) depicts the result of applying the complex dual apodization technique. As it is clear from Fig.~\ref{fig:image_apodization}-(c), the result of the complex dual apodization is by far better than that of the dual apodization. It should also be noted that, the computational complexity of the complex dual apodization is slightly higher than that of the dual apodization.

\subsection{Image De-noising}
\begin{figure}
\centering
\begin{tikzpicture}[yshift=0.00001cm][font=\Large]
  \node (img1)  {\includegraphics[height=6cm,width=8cm]{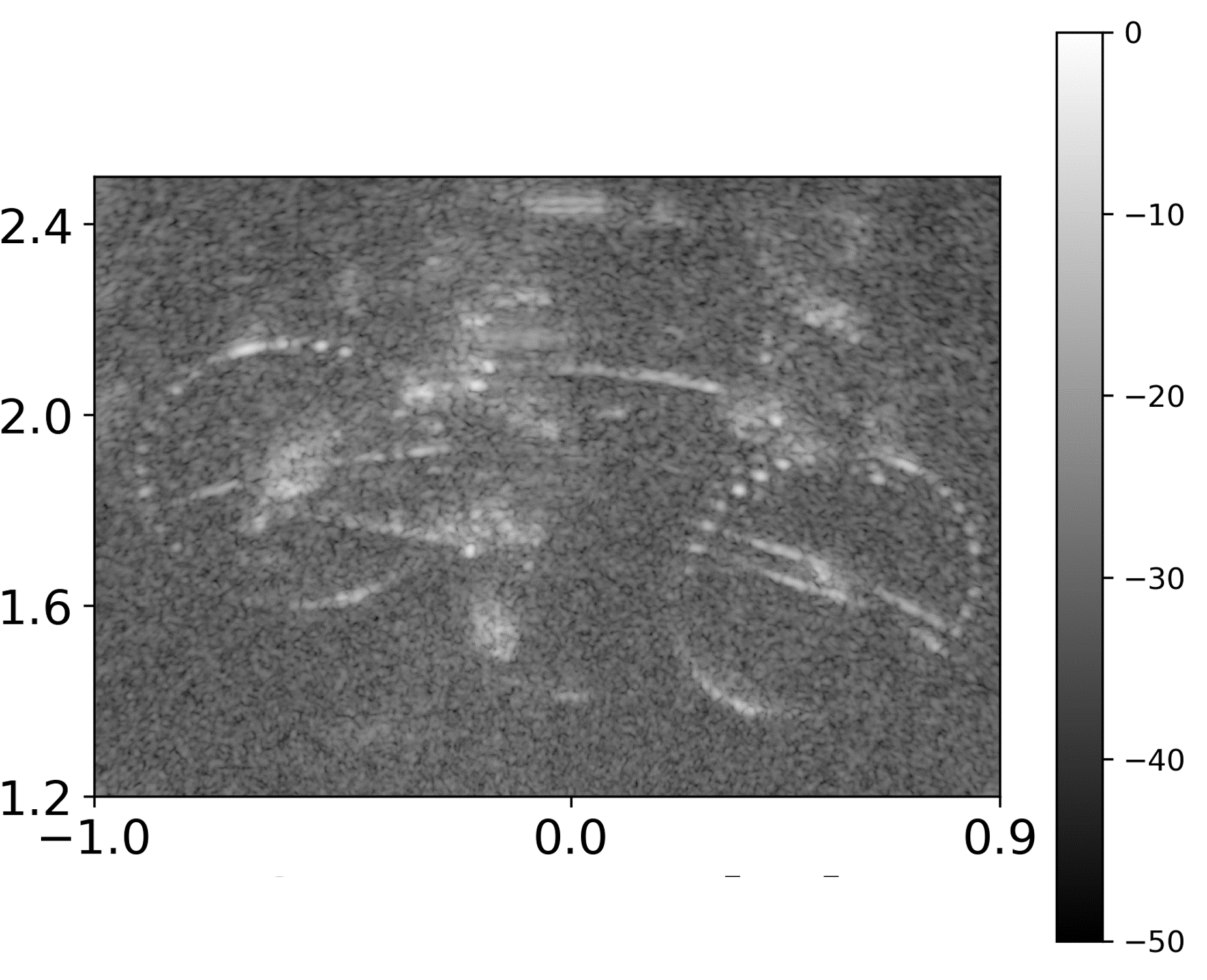}};
   \node[left=of img1, node distance=0cm, rotate = 90, xshift=1cm, yshift=-0.9cm,font=\color{black}] {{Range [m]}};
  \node[below=of img1, node distance=0cm, xshift=0cm, yshift=1.9cm,font=\color{black}] {{Cross Range [m]}};
 \node[below=of img1, node distance=0cm, xshift=0cm, yshift=1.2cm,font=\color{black}] {{(a)}};
\end{tikzpicture}
\begin{tikzpicture}[yshift=0.05cm][font=\Large]
  \node (img1)  {\includegraphics[height=6cm,width=8cm]{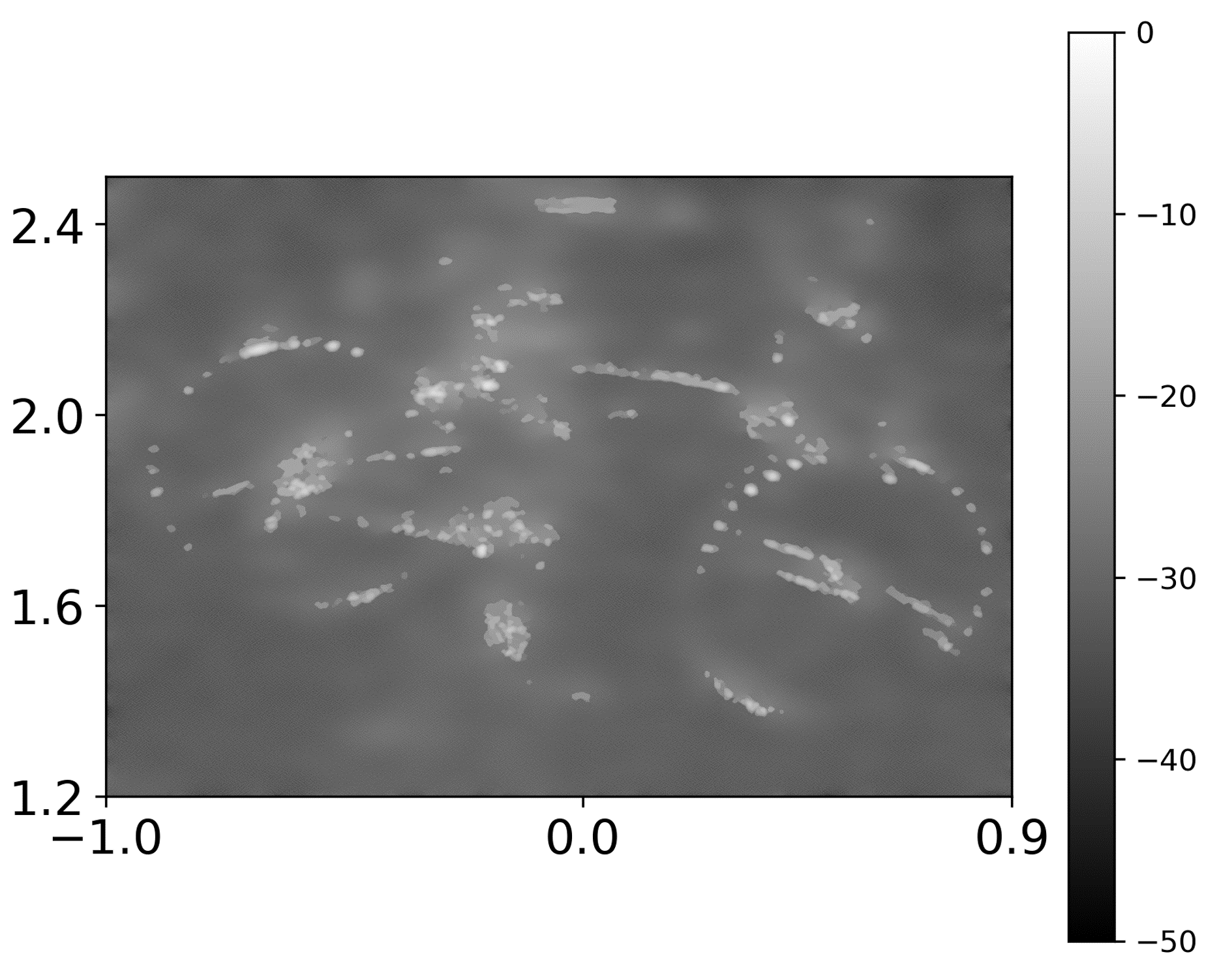}};
 \node[left=of img1, node distance=0cm, rotate = 90, xshift=1cm, yshift=-0.9cm,font=\color{black}] {{Range [m]}};
  \node[below=of img1, node distance=0cm, xshift=0cm, yshift=1.9cm,font=\color{black}] {{Cross Range [m]}};
\node[below=of img1, node distance=0cm, xshift=0cm, yshift=1.2cm,font=\color{black}] {{(b)}};
\end{tikzpicture}
\begin{tikzpicture}[yshift=0.05cm][font=\Large]
  \node (img1)  {\includegraphics[height=6cm,width=8cm]{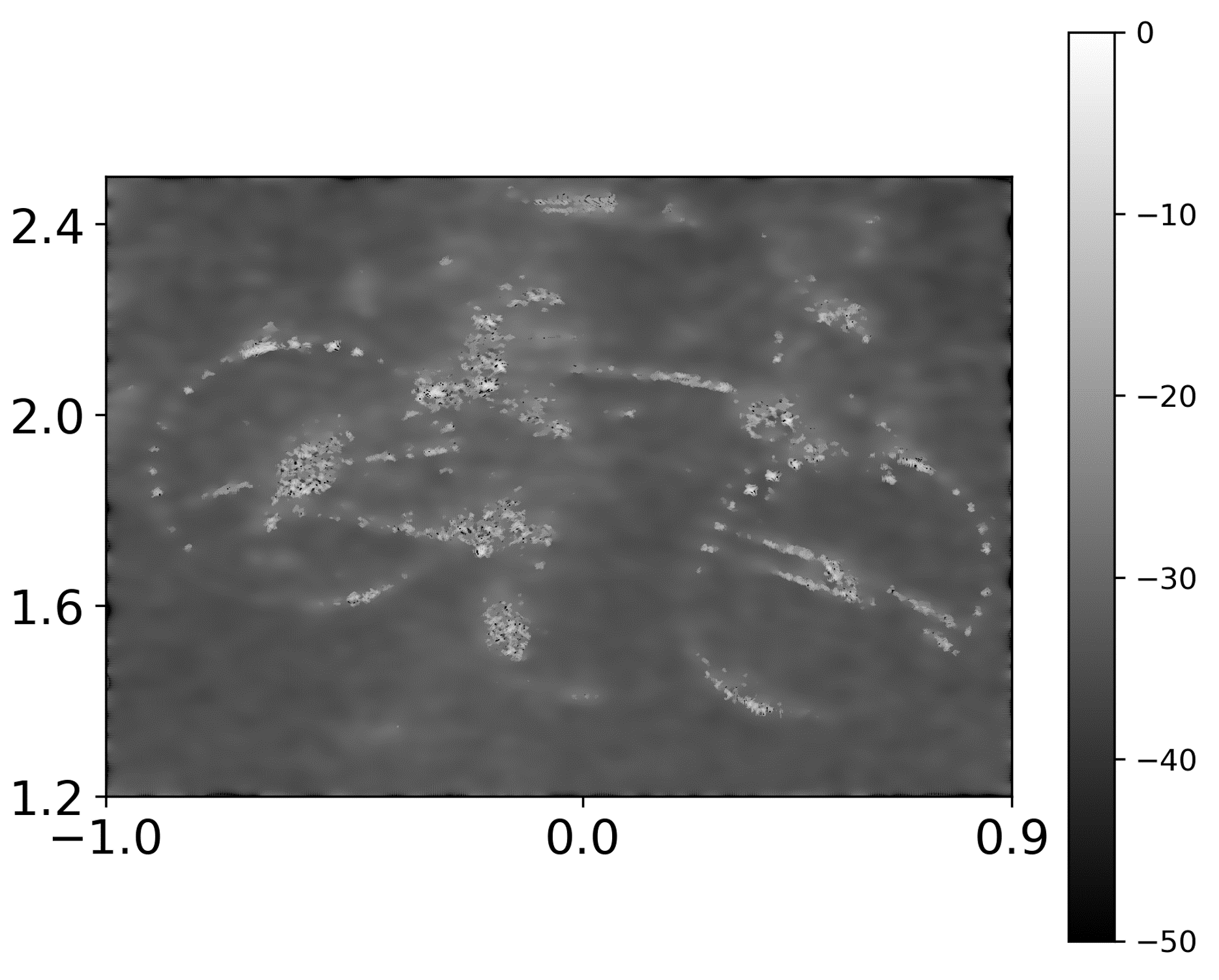}};
 \node[left=of img1, node distance=0cm, rotate = 90, xshift=1cm, yshift=-0.9cm,font=\color{black}] {{Range [m]}};
  \node[below=of img1, node distance=0cm, xshift=0cm, yshift=1.9cm,font=\color{black}] {{Cross Range [m]}};
\node[below=of img1, node distance=0cm, xshift=0cm, yshift=1.2cm,font=\color{black}] {{(c)}};
\end{tikzpicture}

\caption{a) the noisy image, b) the de-noised image, c) the de-noised image by applying the de-noising technique given in (\ref{denoising_opt_EL}) to the result of the complex dual apodization method shown in Fig.~\ref{fig:image_apodization}-(c).}
\label{fig:image_denoising}
\end{figure}

In this sub-section, we implement the modified Perona-Malik method, given in (\ref{denoising_opt_EL}), directly in Python to remove the noise from the reconstructed images. 
The result of the de-noised image has been shown in Fig.~\ref{fig:image_denoising}. In order for the effect of the de-noising process to be seen clearly, we have shown all the images with $\text {50 dB}$ dynamic range.
As can be seen from Fig.~\ref{fig:image_denoising}-(b), the de-noising algorithm provides a remarkable result. The fine structures of the target can be uncovered from the noise while the edges have been preserved. The result shown in  Fig.~\ref{fig:image_denoising}-(b), has been obtained by applying  (\ref{denoising_opt_EL}) to the windowed image shown in Fig.~\ref{fig:image_denoising}-(a). 
In  Fig.~\ref{fig:image_denoising}-(a), we have used Hamming window in both the range and azimuth directions.

In  Fig.~\ref{fig:image_denoising}-(c), we have applied the  de-noising algorithm presented in (\ref{denoising_opt_EL}) to the result of the complex dual apodization which has been shown in Fig.~\ref{fig:image_apodization}-(c). Since the complex dual apodization reduces the side-lobe levels while preserves the resolution limit of the system, therefore,  Fig.~\ref{fig:image_denoising}-(c) provides better results compared to what we have achieved in  Fig.~\ref{fig:image_denoising}-(b). 
In other words, more fine structures related to the bike can be seen in  Fig.~\ref{fig:image_denoising}-(c) compared to what is visible in  Fig.~\ref{fig:image_denoising}-(b).

\subsection{Multi-Spectral-Based Imaging}
\begin{figure}
\centering
\begin{tikzpicture}[yshift=0.00001cm][font=\Large]
  \node (img1)  {\includegraphics[height=6cm,width=8cm]{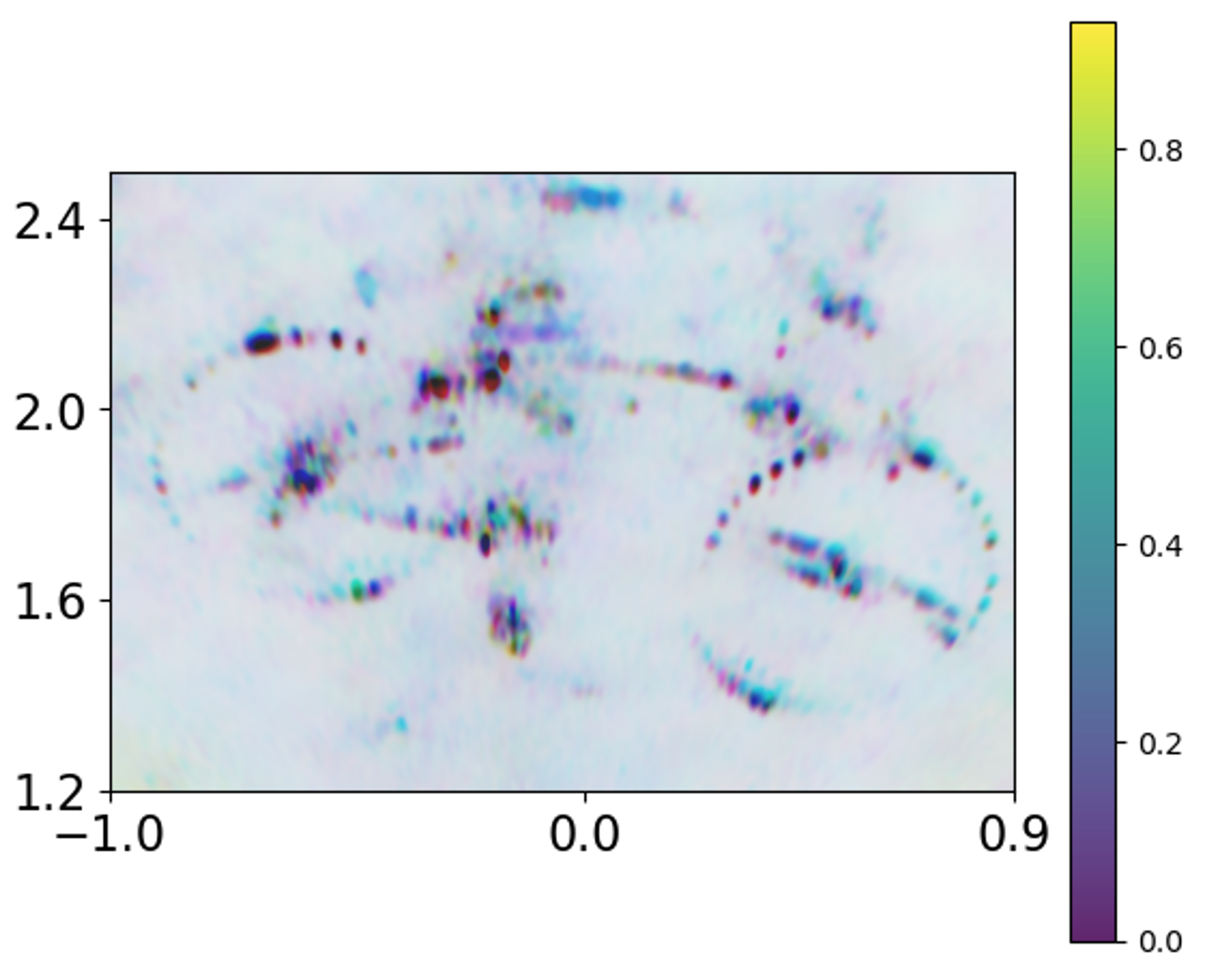}};
   \node[left=of img1, node distance=0cm, rotate = 90, xshift=1cm, yshift=-0.9cm,font=\color{black}] {{Range [m]}};
  \node[below=of img1, node distance=0cm, xshift=0cm, yshift=1.9cm,font=\color{black}] {{Cross Range [m]}};
\end{tikzpicture}
\caption{The result of the multi-spectral-based imaging. The $\text {24 GHz}$ bandwidth has been partitioned into 3 sub-bands.}
\label{fig:multi_spectral_img_bike}
\end{figure}

The result of multi-spectral-based imaging based on 3 sub-bands has been shown in Fig.~\ref{fig:multi_spectral_img_bike}. The result is based on 3 separate images, one image per each sub-band. Then, they are assigned red, green, and blue colors and the final image is created.
In order to increase the visibility of the image, we have performed contrast enhancement as well \cite{img_proc_Gonzalez}.  We have applied the de-noising algorithm given in (\ref{denoising_opt_EL}) to the image prior to partitioning the data into 3 sub-bands.

For better visualization, we have plotted $\rm 1 - |img|$, in which $\rm |img|$ represents the absolute value of the normalized image. Therefore, in Fig.~\ref{fig:multi_spectral_img_bike}, a target with darker color shows the same response over the entire bandwidth of the system. In other words, its radar cross section has less dependency on the range of frequencies that the system is operating over. 

As we mentioned before, the multi-spectral-based imaging can be used for material characterization which will be the subject of our future work.

\subsection{Sub-array Processing}
\begin{figure}
\centering
\begin{tikzpicture}[yshift=0.00001cm][font=\Large]
  \node (img1)  {\includegraphics[height=6cm,width=8cm]{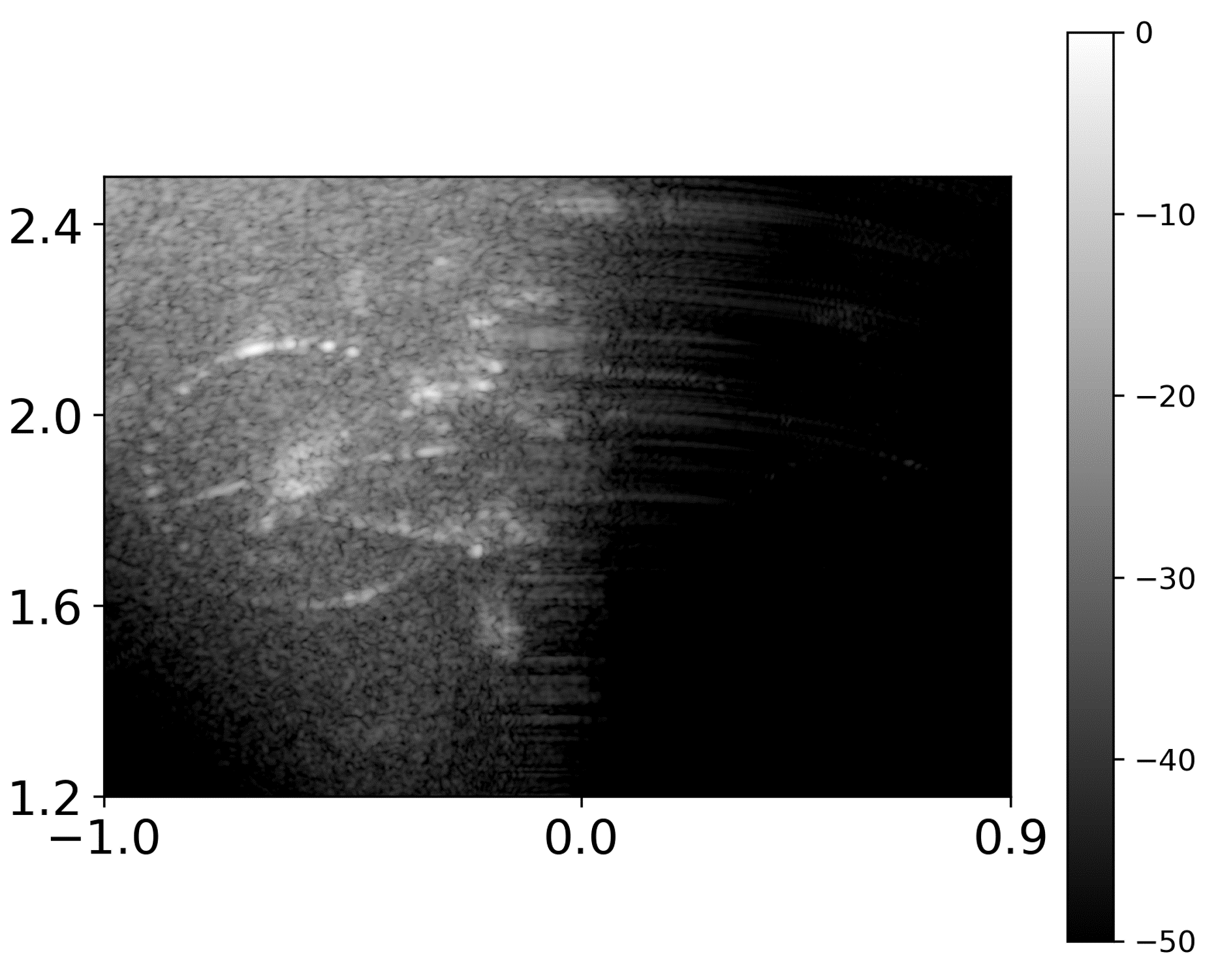}};
   \node[left=of img1, node distance=0cm, rotate = 90, xshift=1cm, yshift=-0.9cm,font=\color{black}] {{Range [m]}};
  \node[below=of img1, node distance=0cm, xshift=0cm, yshift=1.9cm,font=\color{black}] {{Cross Range [m]}};
 \node[below=of img1, node distance=0cm, xshift=0cm, yshift=1.2cm,font=\color{black}] {{(a)}};
\end{tikzpicture}

\begin{tikzpicture}[yshift=0.05cm][font=\Large]
  \node (img1)  {\includegraphics[height=6cm,width=8cm]{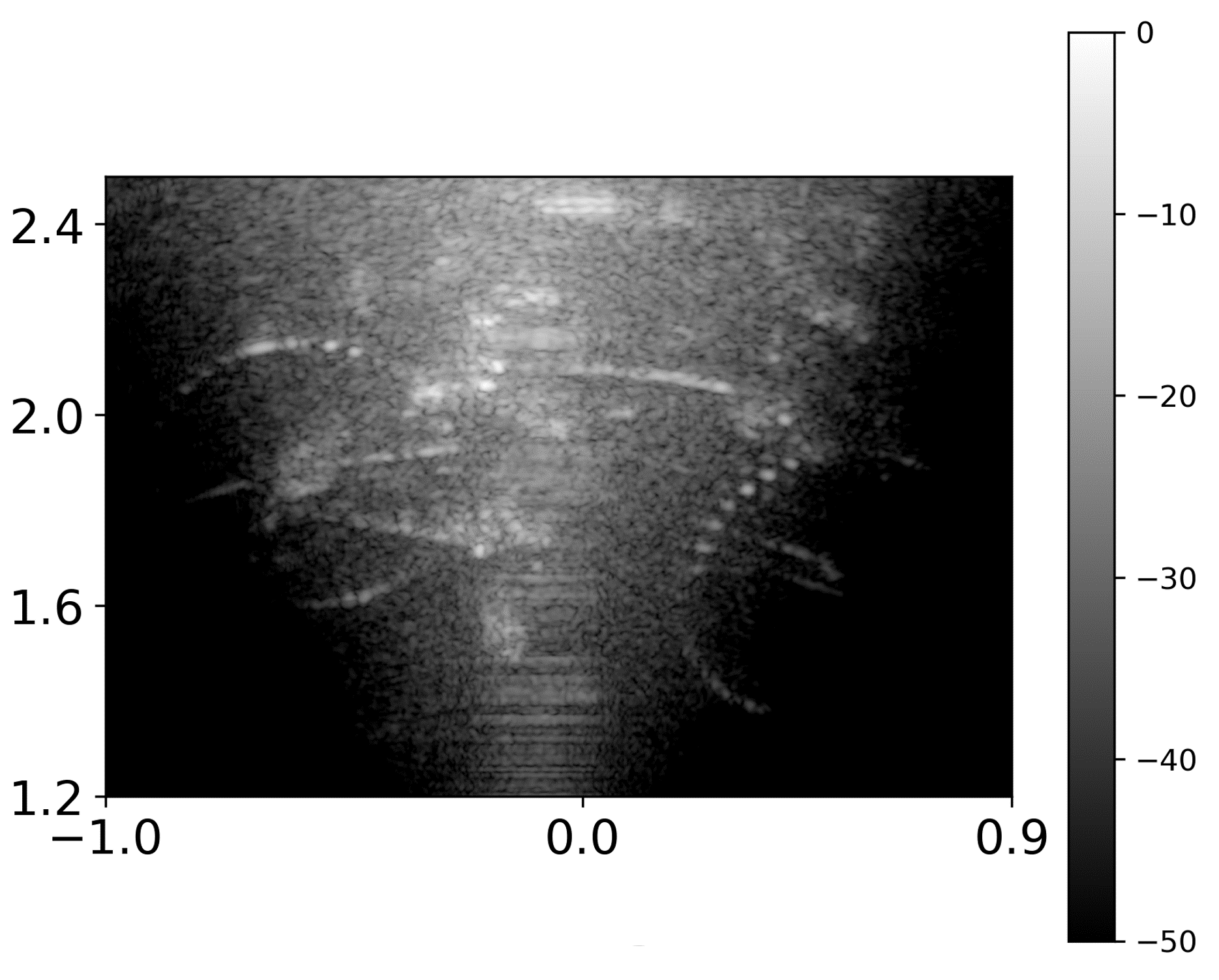}};
 \node[left=of img1, node distance=0cm, rotate = 90, xshift=1cm, yshift=-0.9cm,font=\color{black}] {{Range [m]}};
  \node[below=of img1, node distance=0cm, xshift=0cm, yshift=1.9cm,font=\color{black}] {{Cross Range [m]}};
\node[below=of img1, node distance=0cm, xshift=0cm, yshift=1.2cm,font=\color{black}] {{(b)}};
\end{tikzpicture}

\begin{tikzpicture}[yshift=0.05cm][font=\Large]
  \node (img1)  {\includegraphics[height=6cm,width=8cm]{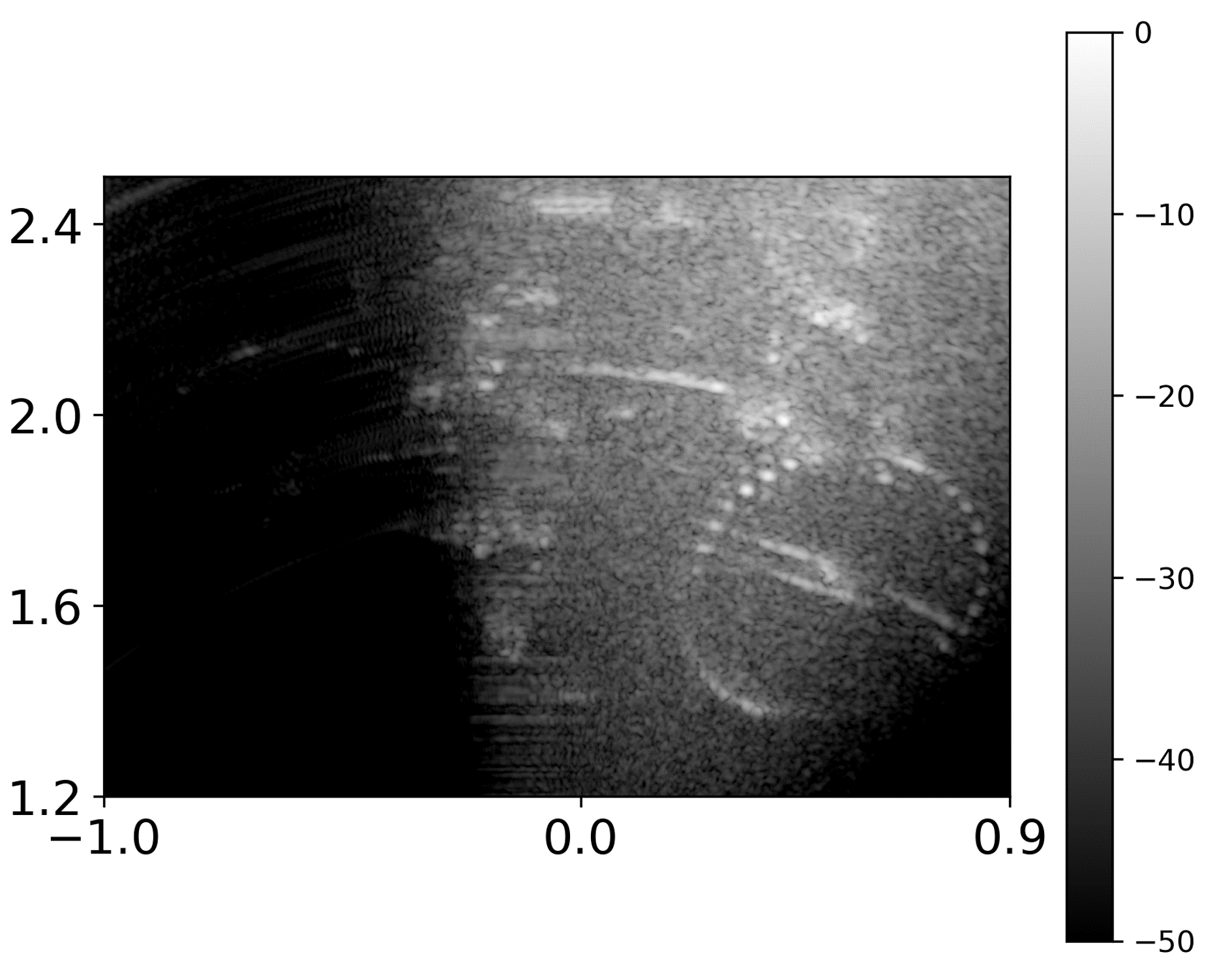}};
 \node[left=of img1, node distance=0cm, rotate = 90, xshift=1cm, yshift=-0.9cm,font=\color{black}] {{Range [m]}};
  \node[below=of img1, node distance=0cm, xshift=0cm, yshift=1.9cm,font=\color{black}] {{Cross Range [m]}};
\node[below=of img1, node distance=0cm, xshift=0cm, yshift=1.2cm,font=\color{black}] {{(c)}};
\end{tikzpicture}
\caption{The result of sub-array-based processing. Sub-arrays of length equal to 10 have been created and all the sub-beams have been directed towards the location of the bike. Finally, all the elements of the array have been used to perform the beam-forming,  a) steering the beam toward left by $\rm 15^o$, b) steering the beam toward the center c) steering the beam toward right by $\rm 15^o$.}
\label{fig:image_sub_array}
\end{figure}

This subsection discusses the sub-array-based processing results.  Fig.~\ref{fig:image_sub_array} shows remarkable results for implementing the sub-array-based processing with {$m = 10$, where $m$ is the length of the sub-arrays.}   
We have partitioned the entire array into sub-arrays with length 10 and with 9 elements overlap between them. Subsequently, we use all these sub-arrays to perform the final stage of the beam-forming process and obtain the image. 
In Fig.~\ref{fig:image_sub_array}-(a), we have steered the beam toward left by $\rm 15^o$. In Fig.~\ref{fig:image_sub_array}-(b), we are illuminating the bore-sight and in Fig.~\ref{fig:image_sub_array}-(c), we have directed the beam toward right by $\rm 15^o$.

By performing sub-array-based processing with sub-arrays of length equal to 10, we can achieve approximately 3 dB gain which is impressive. The reason that we can not obtain 10 dB gain, which is the real number for creating sub-arrays with the length equal to 10, is that when we create the sub-arrays, the overlap between them will force the noise of the system to become correlated. As a result, when we perform the final stage of the beam-forming process to generate the final image, we will be deprived of the full array gain because of the correlation between the noise samples. 
The results that we have presented in Fig.~\ref{fig:image_sub_array}, are based on $\text {50 dB}$ dynamic range and have been presented in gray scale, thus, we can compare them with the reconstructed image shown in Fig.~\ref{fig:image_denoising}-(a). 
The complete analysis of the gain of the sub-array-based processing and its dependency on the length of the sub-arrays, will deviate us from the main subject of this paper and will be the subject of our future work.   
It should be noted that, the gain that we obtain by using the sub-array-based processing technique, which we have presented in this section, is exclusively the gain that is provided by the signal processing unit and that is the reason that the sub-array-based method is considered to be highly valuable. 

However, as we mentioned before and as it is clear from the results presented in Fig.~\ref{fig:image_sub_array}, when we increase the number of elements per each sub-array, the overall field of view of the system becomes narrower. In other words, achieving higher gain using the sub-array-based approach comes at the cost of narrowing the field of view of the imagery system. As a result, in order to be able to perform imaging for the entire field of view of the imagery system, we are required to steer the beam similar to what we have presented in Fig.~\ref{fig:image_sub_array}. 
\begin{figure}
\centering
\begin{tikzpicture}[yshift=0.00001cm][font=\Large]
  \node (img1)  {\includegraphics[height=5cm,width=7cm]{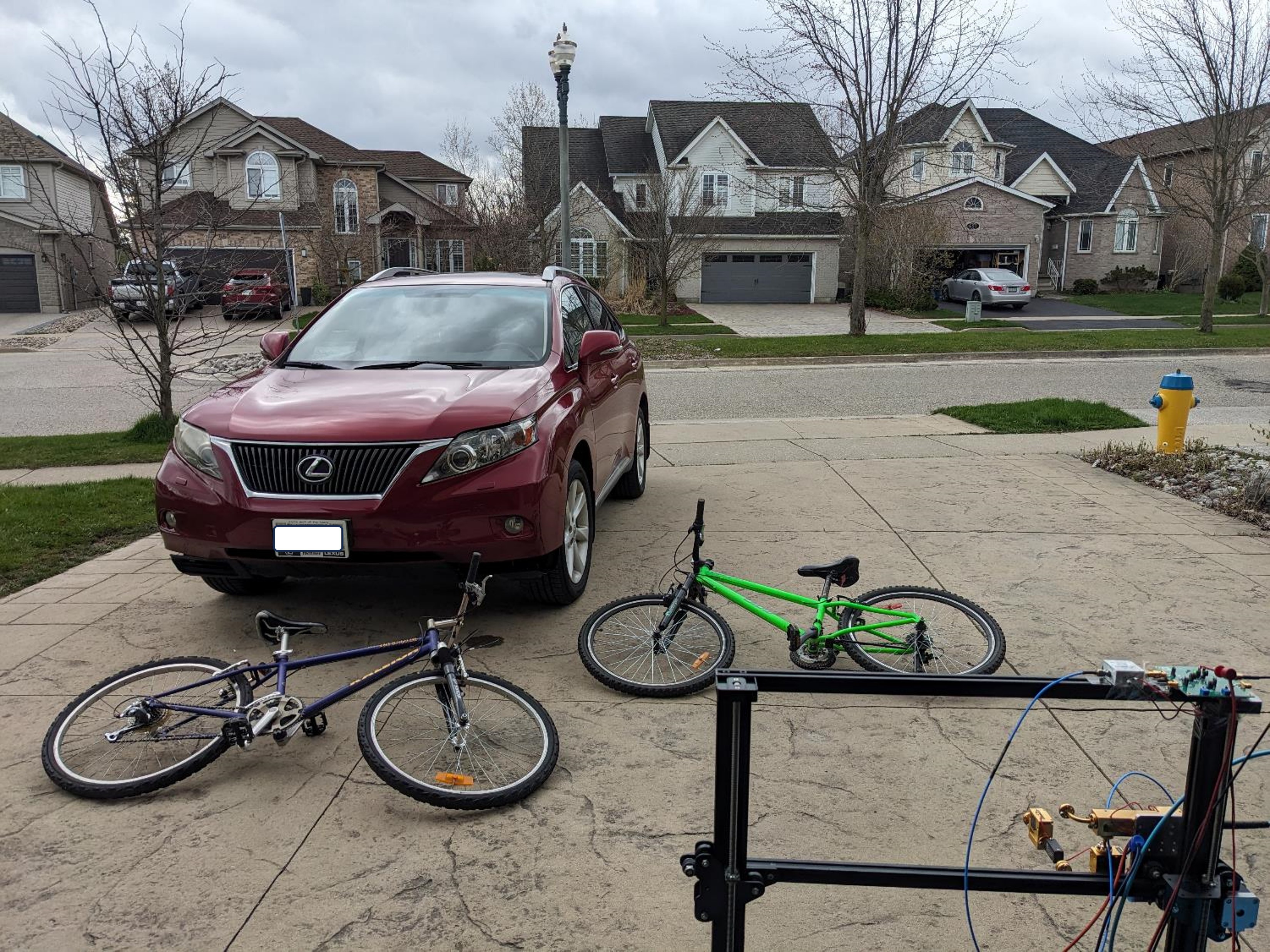}};
\node[below=of img1, node distance=0cm, xshift=-0.5cm, yshift=1cm,font=\color{black}] {{(a)}};
\end{tikzpicture}

\begin{tikzpicture}[yshift=0.05cm][font=\Large]
  \node (img1)  {\includegraphics[height=8cm,width=7cm]{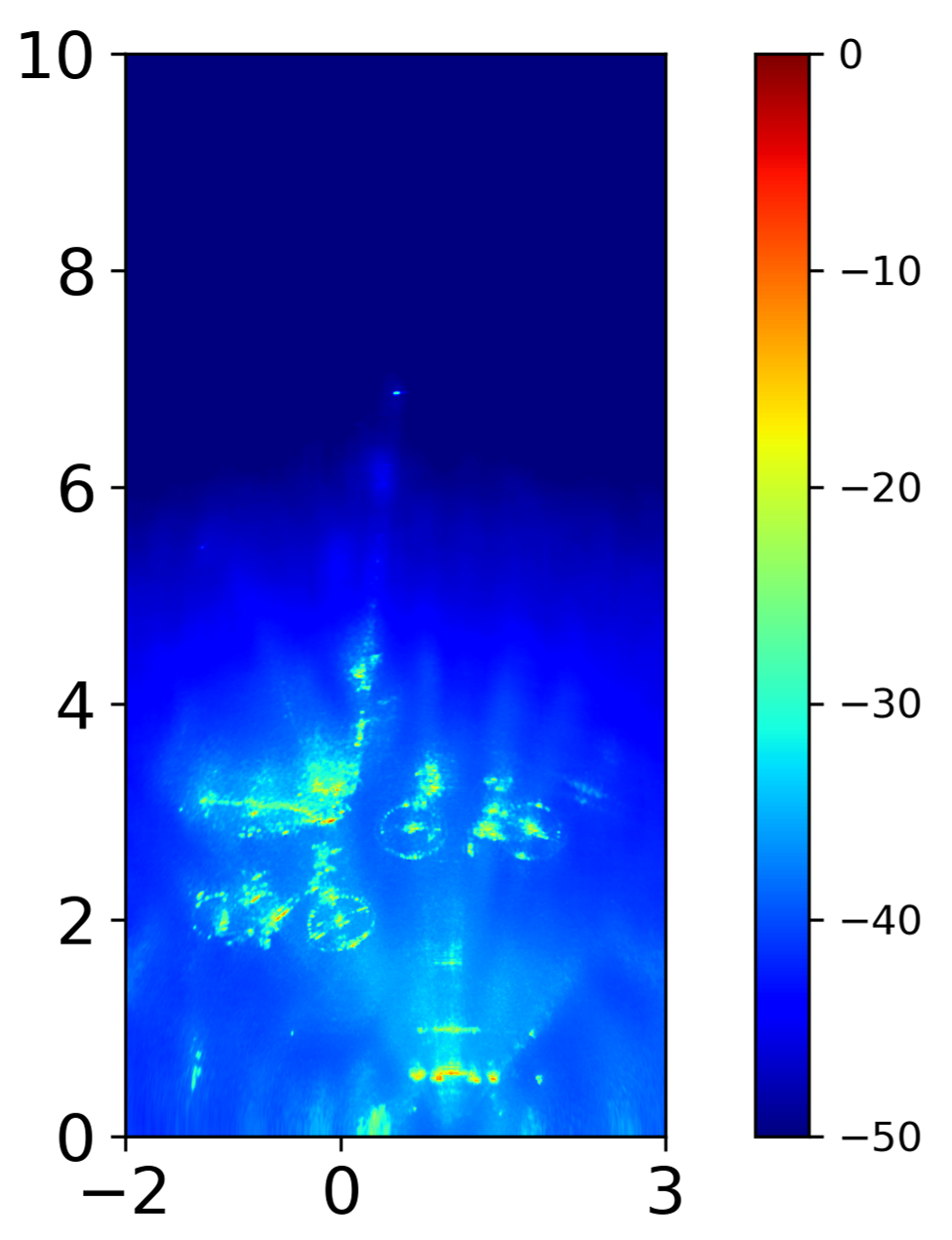}};
 \node[left=of img1, node distance=0cm, rotate = 90, xshift=1.3cm, yshift=-1cm,font=\color{black}] {{Range [m]}};
  \node[below=of img1, node distance=0cm, xshift=-0.5cm, yshift=1.3cm,font=\color{black}] {{Cross Range [m]}};
\node[below=of img1, node distance=0cm, xshift=-0.5cm, yshift=0.6cm,font=\color{black}] {{(b)}};
\end{tikzpicture}

\caption{a) the experimental set-up for a car and two bikes in front of the radar system, b) the reconstructed image.}
\label{fig:car_two_bikes}
\end{figure}

\subsection{More Results}

In this subsection, we present more experimental results from targets such as bike as well as car at near and far distances from the imagery system. It should be noted that, for all the experimental results presented in this subsection, we have used the back-projection algorithm, based on (\ref{BP_algo}), to reconstruct the images. Also, we have used the de-noising algorithm based on (\ref{denoising_opt_EL}) to de-noise the images.

We have performed a test in which a car and two bikes have been located in front of the radar system. Fig.~\ref{fig:car_two_bikes}-(a) illustrates the experimental set-up. In Fig.~\ref{fig:car_two_bikes}-(b), we have presented the reconstructed image. From Fig.~\ref{fig:car_two_bikes}-(b), we can see all the details of all targets that are present in the radar's field of view.

\begin{figure}
\centering
\begin{tikzpicture}[yshift=0.00001cm][font=\Large]
  \node (img1)  {\includegraphics[height=5cm,width=7cm]{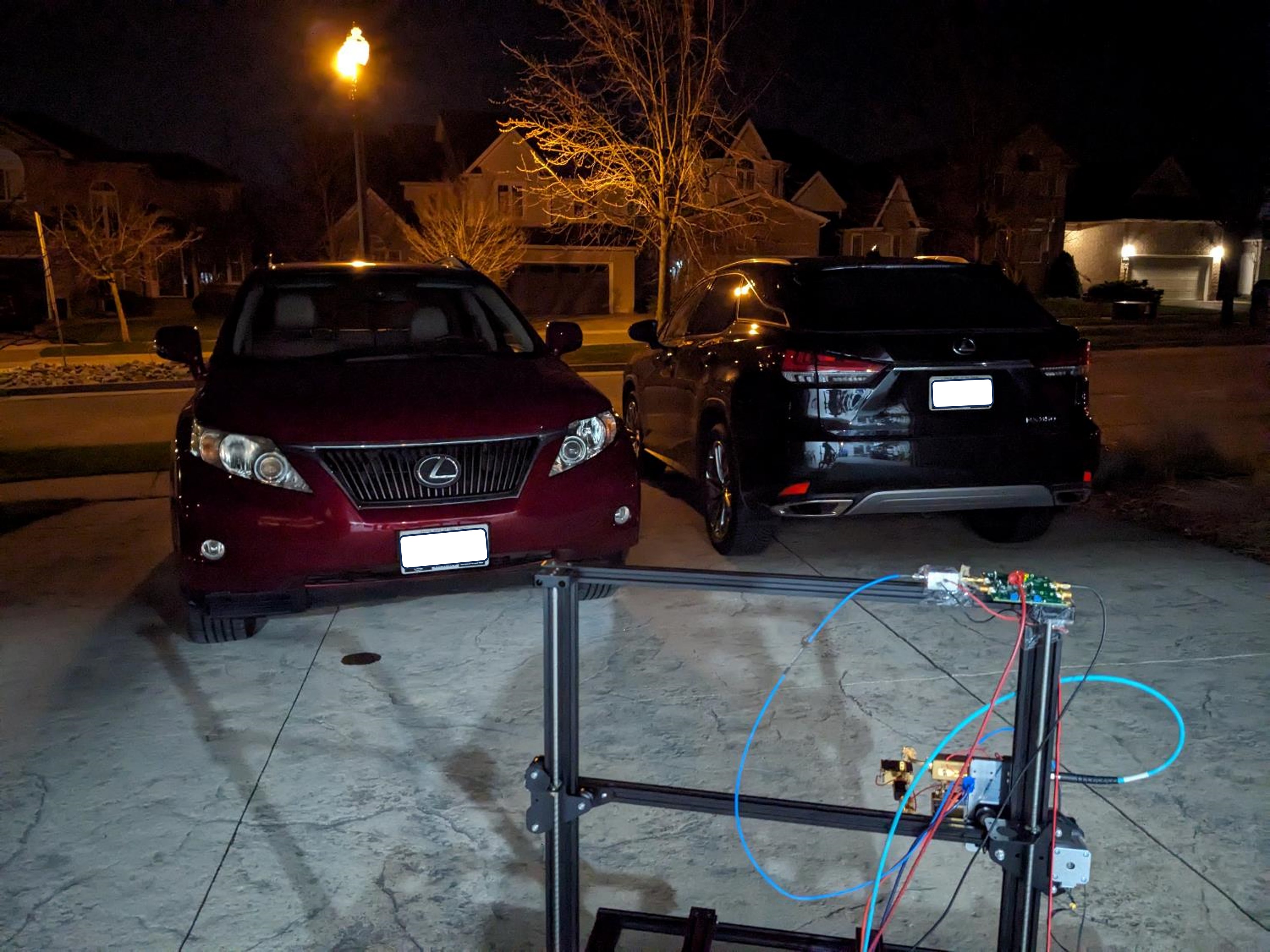}};
\node[below=of img1, node distance=0cm, xshift=-0.5cm, yshift=1cm,font=\color{black}] {{(a)}};
\end{tikzpicture}

\begin{tikzpicture}[yshift=0.05cm][font=\Large]
  \node (img1)  {\includegraphics[height=8cm,width=8cm]{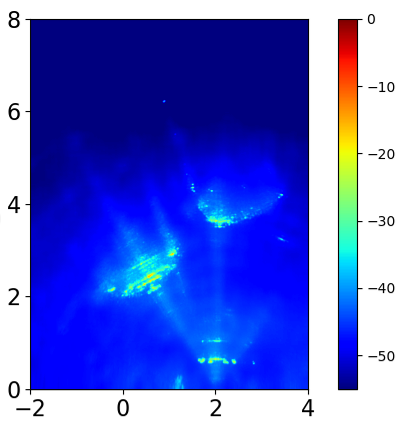}};
 \node[left=of img1, node distance=0cm, rotate = 90, xshift=1.3cm, yshift=-0.8cm,font=\color{black}] {{Range [m]}};
  \node[below=of img1, node distance=0cm, xshift=-0.5cm, yshift=1.3cm,font=\color{black}] {{Cross Range [m]}};
\node[below=of img1, node distance=0cm, xshift=-0.5cm, yshift=0.6cm,font=\color{black}] {{(b)}};
\end{tikzpicture}

\caption{a) the experimental set-up for 2 cars, the data has been collected during night time, b) the reconstructed image.}
\label{fig:2_cars}
\end{figure}

We have conducted a different experiment which includes 2 cars in front of the imagery system. Fig.~\ref{fig:2_cars}-(a) illustrates the experimental set-up. As can be seen from Fig.~\ref{fig:2_cars}-(a) , the data has been collected during night time. 
In Fig.~\ref{fig:2_cars}-(b), we have shown the reconstructed image. From Fig.~\ref{fig:2_cars}-(b), we can see the details of both cars such as the bumper, headlights, license plates, and side mirrors.

\begin{figure}
\centering
\begin{tikzpicture}[yshift=0.00001cm][font=\Large]
  \node (img1)  {\includegraphics[height=5cm,width=7cm]{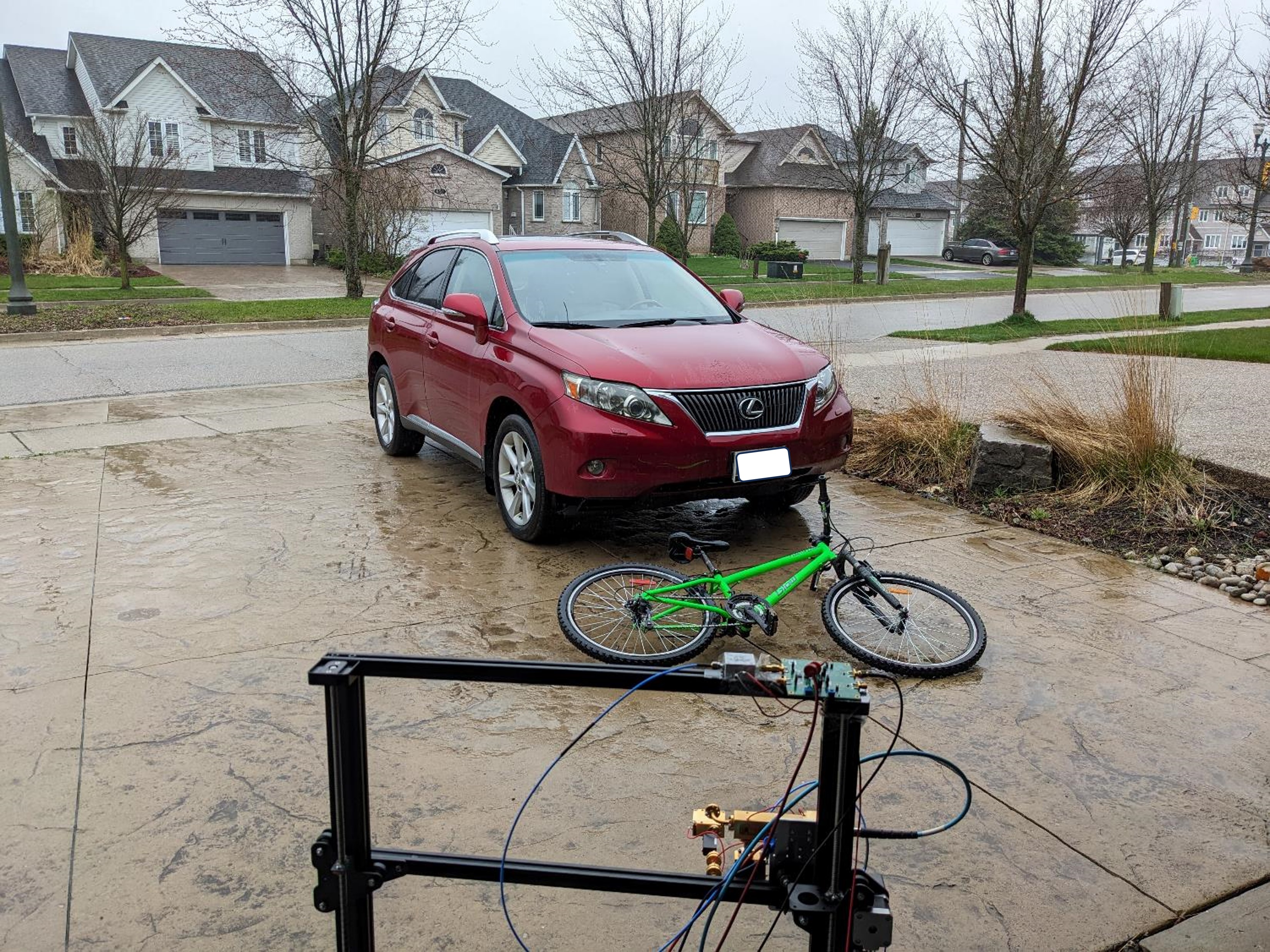}};
\node[below=of img1, node distance=0cm, xshift=-0.5cm, yshift=1cm,font=\color{black}] {{(a)}};
\end{tikzpicture}

\begin{tikzpicture}[yshift=0.05cm][font=\Large]
  \node (img1)  {\includegraphics[height=6cm,width=6cm]{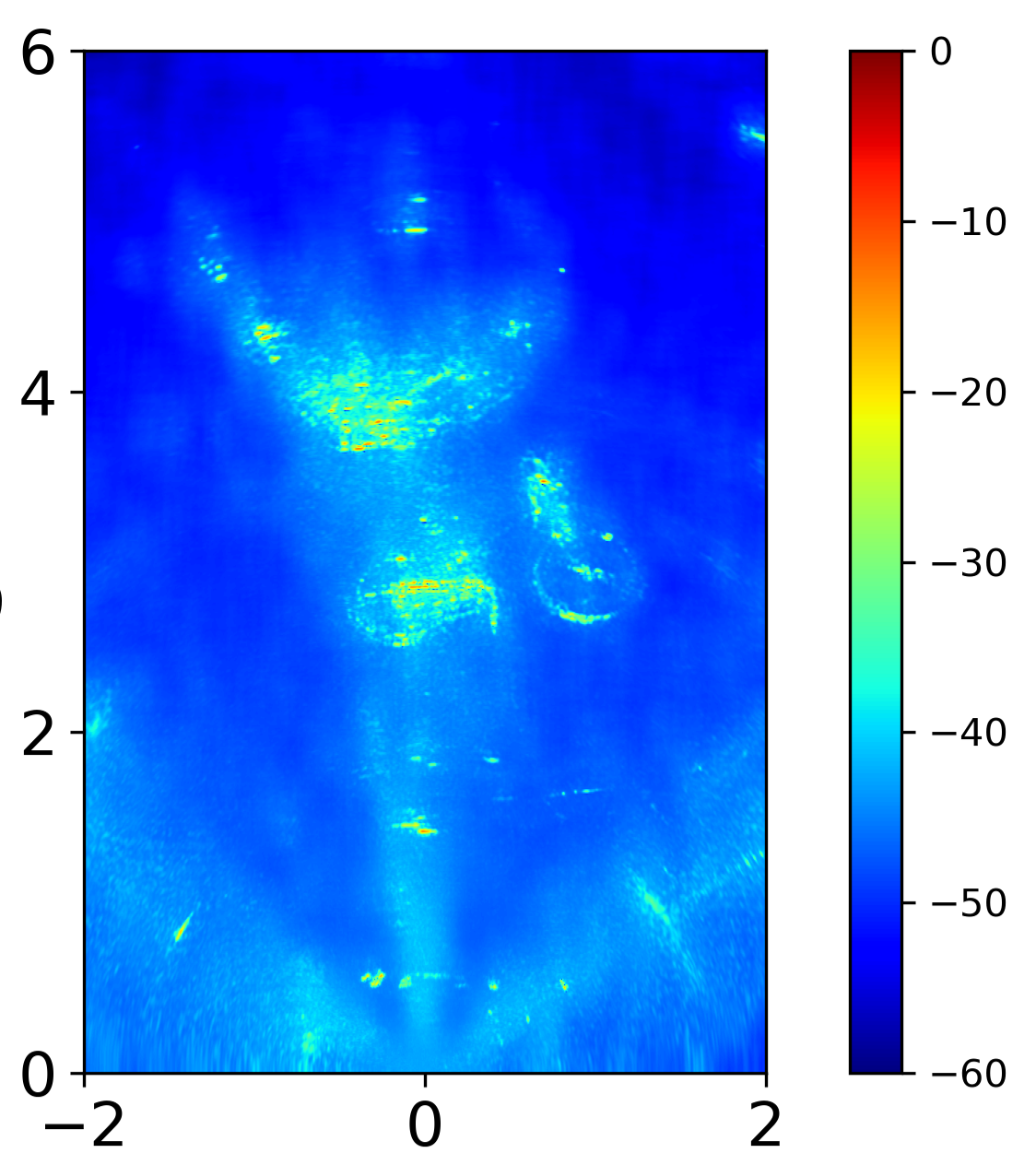}};
 \node[left=of img1, node distance=0cm, rotate = 90, xshift=1.3cm, yshift=-0.8cm,font=\color{black}] {{Range [m]}};
  \node[below=of img1, node distance=0cm, xshift=-0.5cm, yshift=1.3cm,font=\color{black}] {{Cross Range [m]}};
\node[below=of img1, node distance=0cm, xshift=-0.5cm, yshift=0.6cm,font=\color{black}] {{(b)}};
\end{tikzpicture}
\caption{a) the experimental set-up for the bike in front of the car, the data has been collected in a rainy morning, b) the reconstructed image.}
\label{fig:bike_car_rain}
\end{figure}

We have conducted an additional test. The data for this scenario has been collected in rainy weather. Fig.~\ref{fig:bike_car_rain}-(a) illustrates the experimental set-up for the bike situated on the ground in front of the car. As we mentioned, the data has been collected in a rainy morning in order to test the ability of the imagery system to create high resolution images from the targets even in rainy conditions.  As can be seen from Fig.~\ref{fig:bike_car_rain}-(a), the ground, the bike, and the car are all wet and rain is coming during the period that the data is being collected. Fig.~\ref{fig:bike_car_rain}-(b) shows the reconstructed image. 

\begin{figure}
\centering
\begin{tikzpicture}[yshift=0.00001cm][font=\Large]
  \node (img1)  {\includegraphics[height=5cm,width=7cm]{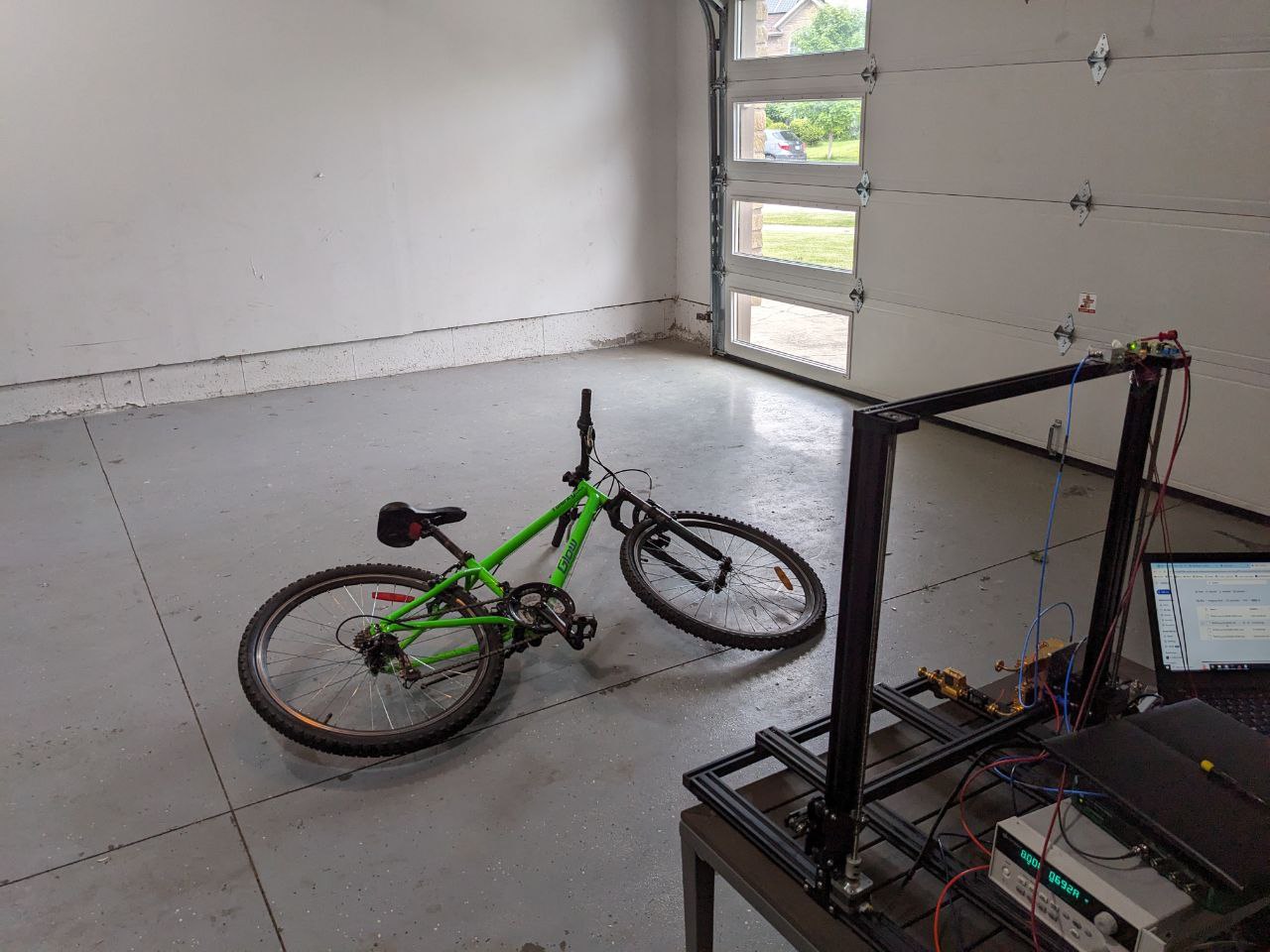}};
\node[below=of img1, node distance=0cm, xshift=-0.5cm, yshift=1cm,font=\color{black}] {{(a)}};
\end{tikzpicture}

\begin{tikzpicture}[yshift=0.05cm][font=\Large]
  \node (img1)  {\includegraphics[height=5cm,width=6cm]{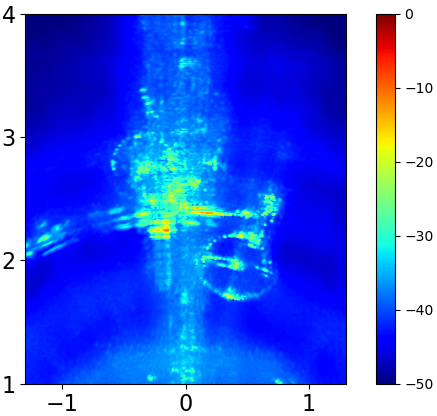}};
 \node[left=of img1, node distance=0cm, rotate = 90, xshift=1.3cm, yshift=-0.8cm,font=\color{black}] {{Range [m]}};
  \node[below=of img1, node distance=0cm, xshift=-0.5cm, yshift=1.3cm,font=\color{black}] {{Cross Range [m]}};
\node[below=of img1, node distance=0cm, xshift=-0.5cm, yshift=0.6cm,font=\color{black}] {{(b)}};
\end{tikzpicture}
\caption{a) the experimental set-up for the bike in front of the imagery system, b) the reconstructed image.}
\label{fig:img_NOFOG}
\end{figure}

\begin{figure}
\centering
\begin{tikzpicture}[yshift=0.00001cm][font=\Large]
  \node (img1)  {\includegraphics[height=5cm,width=7cm]{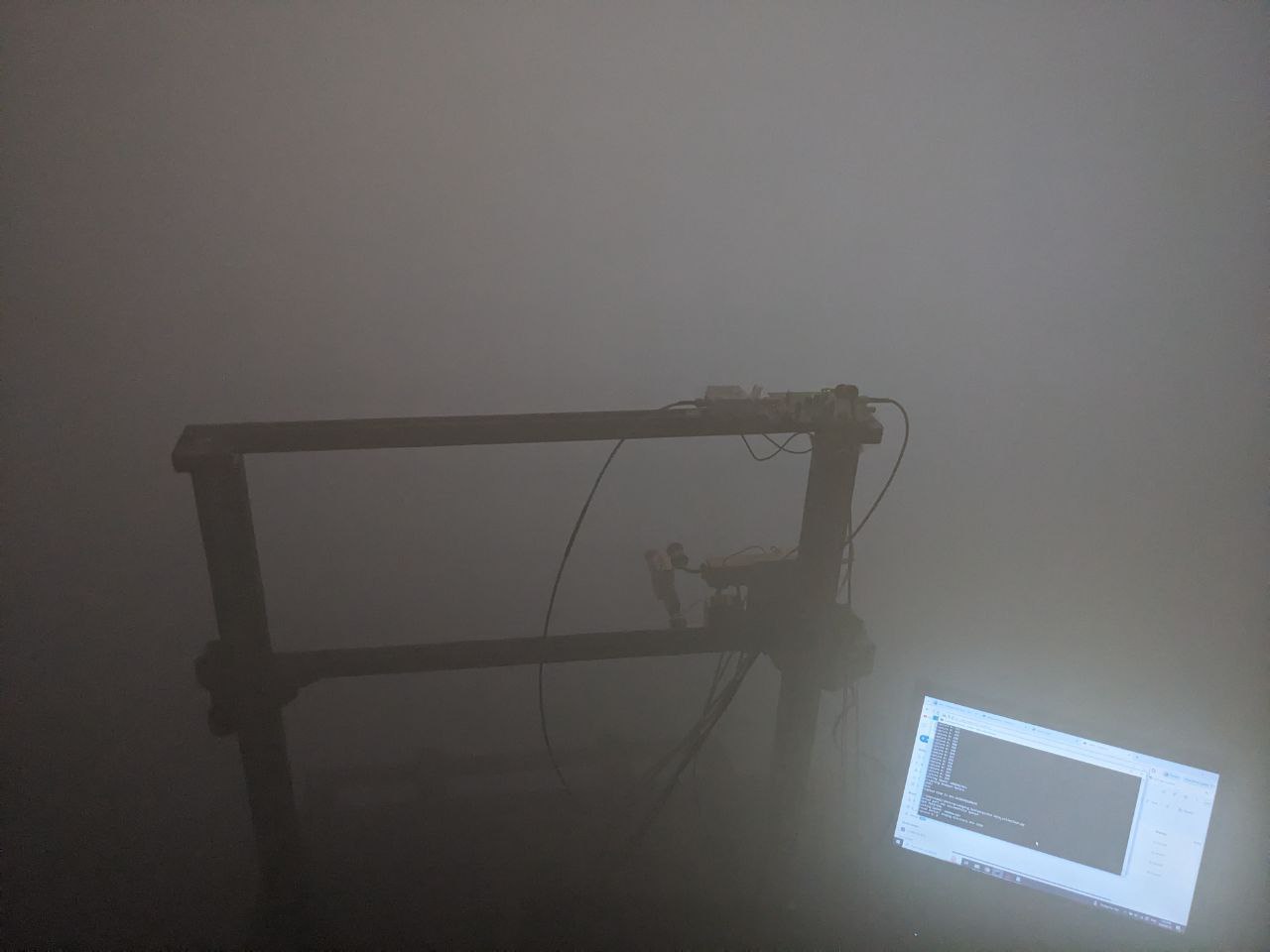}};
\node[below=of img1, node distance=0cm, xshift=-0.5cm, yshift=1cm,font=\color{black}] {{(a)}};
\end{tikzpicture}

\begin{tikzpicture}[yshift=0.05cm][font=\Large]
  \node (img1)  {\includegraphics[height=5cm,width=6cm]{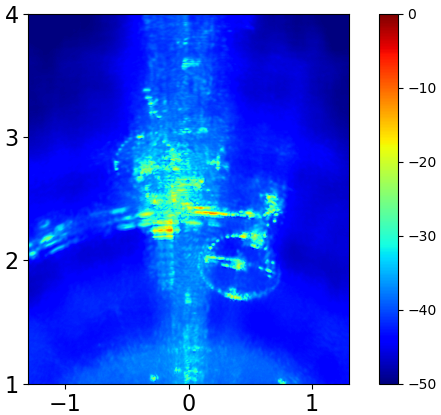}};
 \node[left=of img1, node distance=0cm, rotate = 90, xshift=1.3cm, yshift=-0.8cm,font=\color{black}] {{Range [m]}};
  \node[below=of img1, node distance=0cm, xshift=-0.5cm, yshift=1.3cm,font=\color{black}] {{Cross Range [m]}};
\node[below=of img1, node distance=0cm, xshift=-0.5cm, yshift=0.6cm,font=\color{black}] {{(b)}};
\end{tikzpicture}
\caption{a) the experimental set-up presented in Fig.~\ref{fig:img_NOFOG}-(a), the data has been collected in a foggy condition, b) the reconstructed image.}
\label{fig:img_FOG}
\end{figure}

{To further demonstrate the all-weather capability of the imagery system, we have conducted a test in a foggy environment. The data for this scenario has been collected in dense fog. Fig.~\ref{fig:img_NOFOG}-(a) shows the experimental set-up for the bike situated on the ground in front of the imagery system in clear weather condition and Fig.~\ref{fig:img_NOFOG}-(b) presents the reconstructed image from the scene. The set-up is inside the garage and we have then created a thick fogy situation. In Fig.~\ref{fig:img_FOG}-(a), the test set-up in the fog has been shown. The thick layer of fog makes it impossible for the bike to be seen in the optical image which has been presented in Fig.~\ref{fig:img_FOG}-(a). In Fig.~\ref{fig:img_FOG}-(b), we have illustrated the reconstructed image from the fogy test set-up. It is clear from Fig.~\ref{fig:img_FOG}-(b) that even in thick fogy situation the imagery system is capable of producing high resolution images from the targets. In fact, the reconstructed images under no-fog and foggy conditions, presented in Fig.~\ref{fig:img_NOFOG}-(b) and Fig.~\ref{fig:img_FOG}-(b), respectively,  are identical.}


\begin{figure}
\centering
\begin{tikzpicture}[yshift=0.00001cm][font=\Large]
  \node (img1)  {\includegraphics[height=5cm,width=7cm]{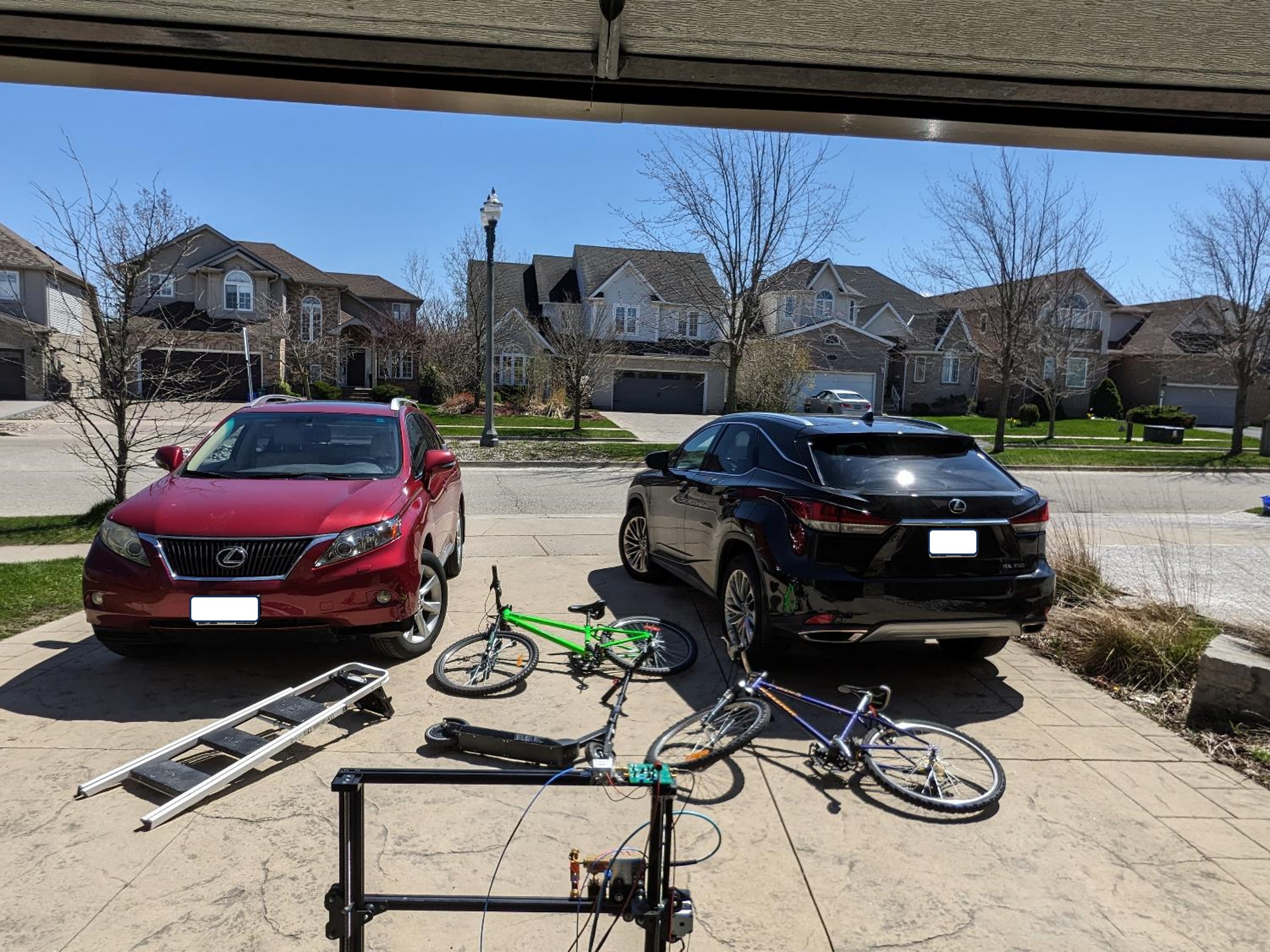}};
\node[below=of img1, node distance=0cm, xshift=-0.5cm, yshift=1cm,font=\color{black}] {{(a)}};
\end{tikzpicture}

\begin{tikzpicture}[yshift=0.05cm][font=\Large]
  \node (img1)  {\includegraphics[height=7cm,width=8cm]{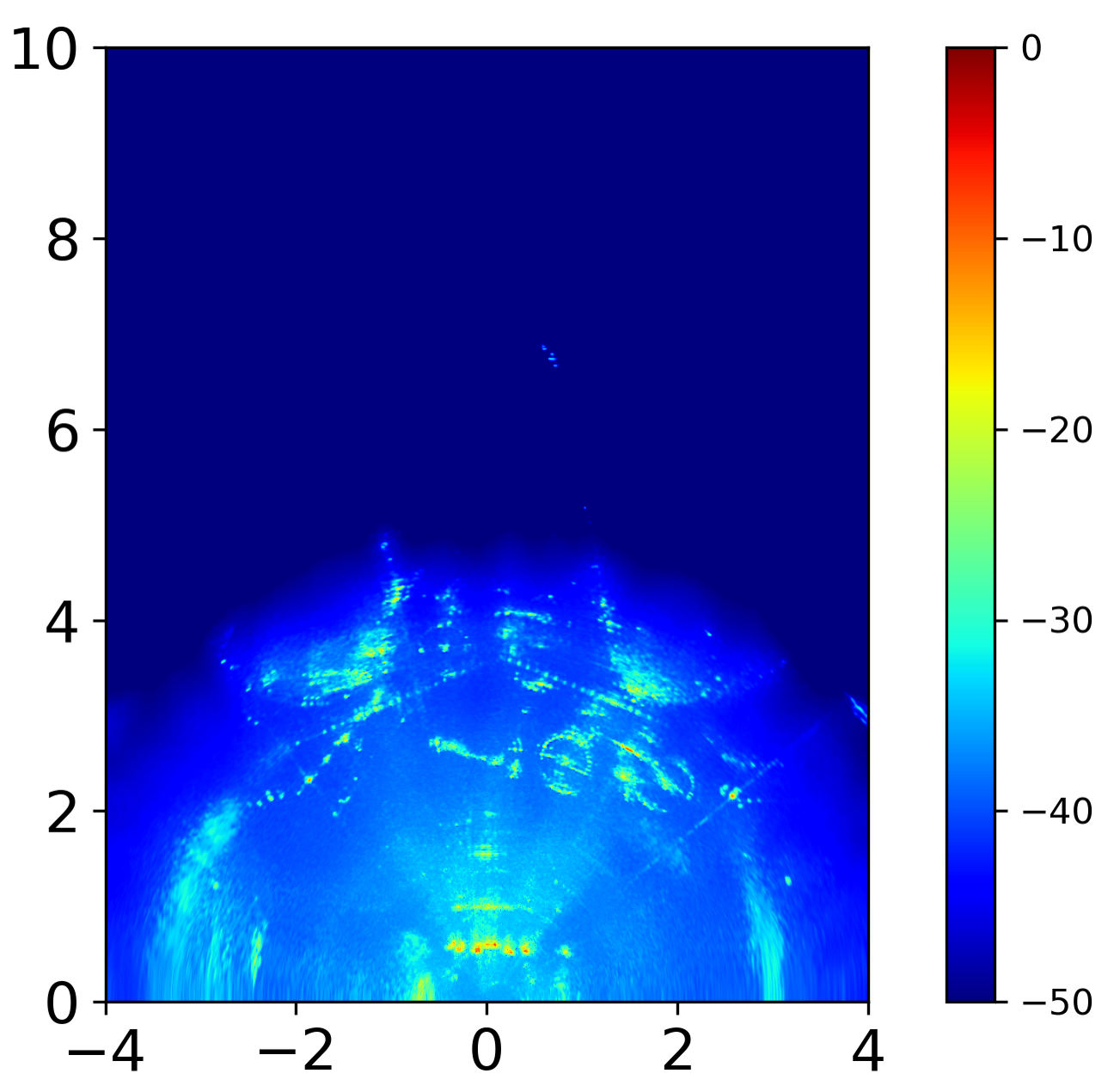}};
 \node[left=of img1, node distance=0cm, rotate = 90, xshift=1.3cm, yshift=-1cm,font=\color{black}] {{Range [m]}};
  \node[below=of img1, node distance=0cm, xshift=-0.5cm, yshift=1.3cm,font=\color{black}] {{Cross Range [m]}};
\node[below=of img1, node distance=0cm, xshift=-0.5cm, yshift=0.7cm,font=\color{black}] {{(b)}};
\end{tikzpicture}

\caption{a) the experimental set-up for 2 cars, 2 bikes, a ladder, and a scooter, b)  the reconstructed image.}
\label{fig:all_targets}
\end{figure}

The next test experiment, which has been shown in Fig.~\ref{fig:all_targets}-(a), includes 2 cars, 2 bikes, a ladder, and a scooter. Fig.~\ref{fig:all_targets}-(b) illustrates the reconstructed image which clearly shows all the targets. The reason we have conducted this experiment is that the number of targets are higher and the targets have been located in close proximity of one another. Consequently, we can see some undesired effects that have been created by multi-path.

\begin{figure}
\centering
\begin{tikzpicture}[yshift=0.00001cm][font=\Large]
  \node (img1)  {\includegraphics[height=5cm,width=7cm]{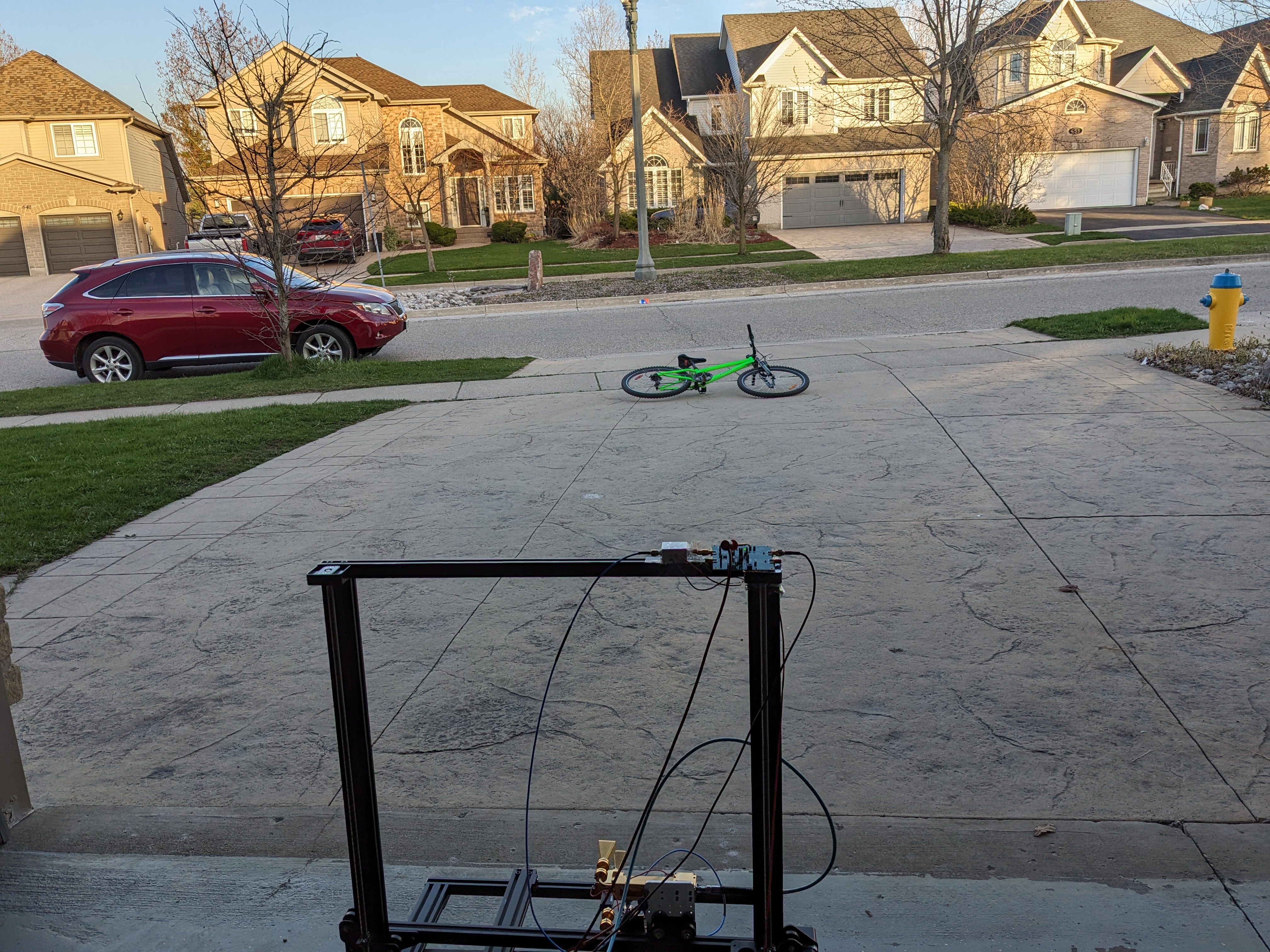}};
\node[below=of img1, node distance=0cm, xshift=0cm, yshift=1cm,font=\color{black}] {{(a)}};
\end{tikzpicture}

\begin{tikzpicture}[yshift=0.05cm][font=\Large]
  \node (img1)  {\includegraphics[height=7cm,width=7cm]{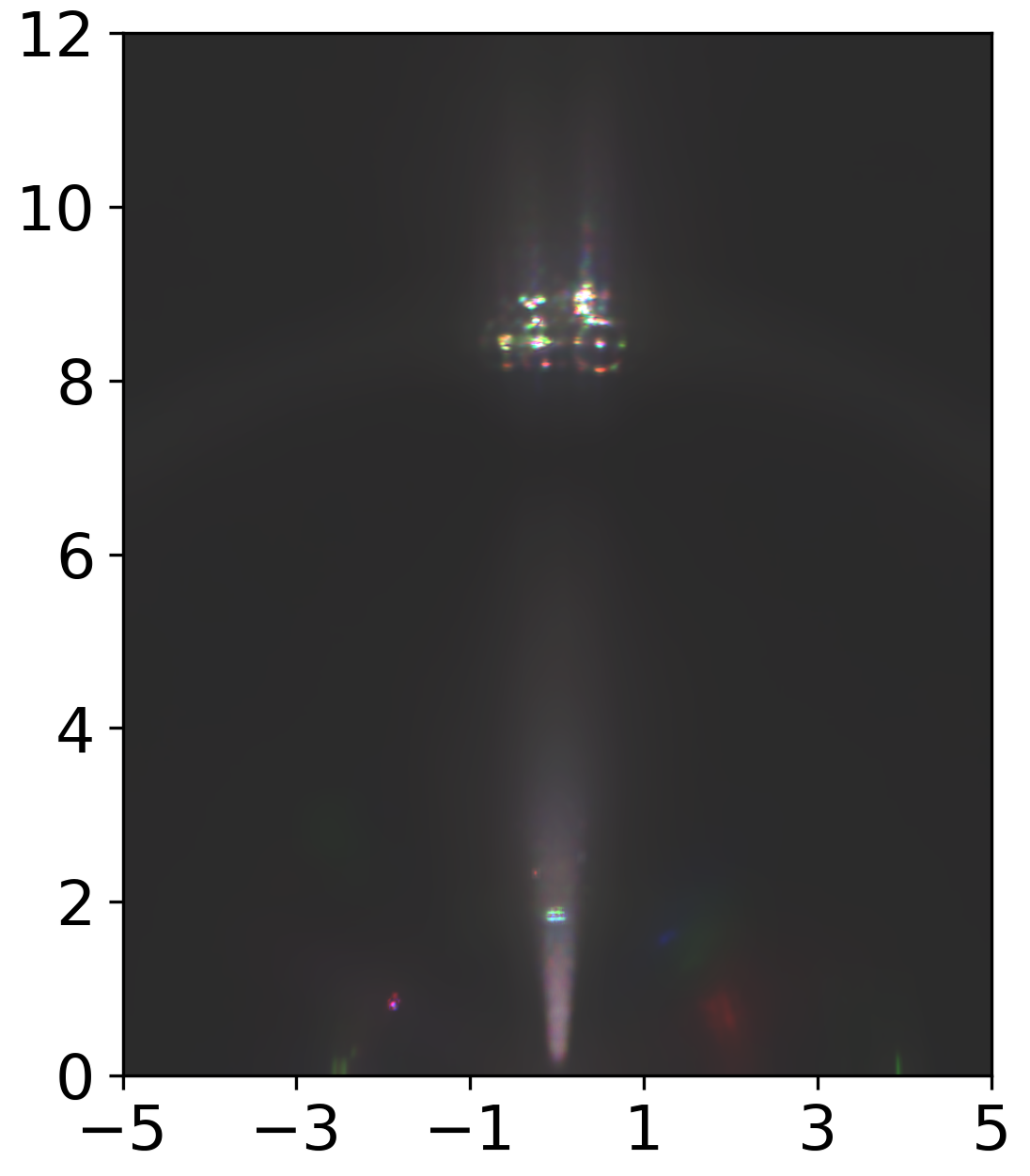}};
 \node[left=of img1, node distance=0cm, rotate = 90, xshift=1.2cm, yshift=-1cm,font=\color{black}] {{Range [m]}};
  \node[below=of img1, node distance=0cm, xshift=0cm, yshift=1.3cm,font=\color{black}] {{Cross Range [m]}};
\node[below=of img1, node distance=0cm, xshift=0cm, yshift=0.6cm,font=\color{black}] {{(b)}};
\end{tikzpicture}

\caption{a) the bike in front of the imagery system at far distance, b) the reconstructed image.}
\label{fig:bike_horn}
\end{figure}

Finally, we conclude this section with the experimental set-up for the bike located at 8.5 m distance in front of the imagery system. 
Fig.~\ref{fig:bike_horn}-(a) depicts the experimental result and in Fig.~\ref{fig:bike_horn}-(b) we have shown the reconstructed image.
When we place the targets at far distances from the imagery system, the SNR as well as the resolution limit of the system in the cross range direction decrease.

It is worth mentioning that, for this experiment we have used horn antennas on both the transmit and receive sides in order to increase the maximum range of the imagery system. Consequently, the field of view of the imagery system has been reduced and in the reconstructed image, which has been shown in Fig.~\ref{fig:bike_horn}-(b), we can only see the bike. 

\section{Conclusion}
In this paper, we presented our wide-band custom-made FMCW radar system which operates at millimeter wave frequencies. We described the functionalities of the system in detail. We further, presented the imaging algorithm and several signal processing techniques for image formation and image post processing. 
At the end, we presented all the results based on the experimental data gathered from the imagery system. 

The paper is a proof that images with high resolution and high quality can be generated using compact, lightweight, and low-cost millimeter wave systems such as the one presented in this work. 

%\section*{Acknowledgment}

\end{document}